\documentclass[useAMS,usenatbib]{mn2e}                                                             
\usepackage[totalwidth=515pt,totalheight=660pt,left=1.4cm,right=1.4cm]{geometry}              
\usepackage{graphicx,amssymb,color}                                                         
\usepackage[normalem]{ulem}                                                                    
\input psfig 

\title[Radiative and drag forces in post-MS systems]
{The orbital evolution of asteroids, pebbles and planets from giant branch stellar radiation and winds}
\author[Veras, Eggl, \& G\"{a}nsicke]{
Dimitri Veras$^{1}$\thanks{E-mail:d.veras@warwick.ac.uk},
Siegfried Eggl$^{2}$,
Boris T. G\"{a}nsicke$^{1}$
\\
$^{1}$Department of Physics, University of Warwick, Coventry CV4 7AL, UK
\\
$^{2}$IMCCE Observatroire de Paris, Univ. Lille 1, UPMC, 77 Av. Denfert-Rochereau, 75014 Paris, France
}
\begin{document}

\date{Accepted 2015 May 07. Received 2015 May 04; in original form 2015 January 28}

\pagerange{\pageref{firstpage}--\pageref{lastpage}} \pubyear{2015} 

\maketitle

\label{firstpage}

\begin{abstract}
The discovery of over 50 planets around evolved stars and more than 35 debris discs orbiting white dwarfs highlight the increasing need to understand small body evolution around both early and asymptotic giant branch (GB) stars.  Pebbles and asteroids are susceptible to strong accelerations from the intense luminosity and winds of GB stars.  Here, we establish equations that can model time-varying GB stellar radiation, wind drag and mass loss.  We derive the complete three-dimensional equations of motion in orbital elements due to (1) the Epstein and Stokes regimes of stellar wind drag, (2) Poynting-Robertson drag, and (3) the Yarkovsky drift with seasonal and diurnal components.  We prove through averaging that the potential secular eccentricity and inclination excitation due to Yarkovsky drift can exceed that from Poynting-Robertson drag and radiation pressure by at least three orders of magnitude, possibly flinging asteroids which survive YORP spin-up into a widely dispersed cloud around the resulting white dwarf.  The GB Yarkovsky effect alone may change an asteroid's orbital eccentricity by ten per cent in just one Myr.  Damping perturbations from stellar wind drag can be just as extreme, but are strongly dependent on the highly uncertain local gas density and mean free path length.  We conclude that GB radiative and wind effects must be considered when modelling the post-main-sequence evolution of bodies smaller than about 1000 km.
\end{abstract}


\section{Introduction}
Robust observational evidence for exoplanetary systems around both main sequence (MS) and 
post-MS stars demands a self-consistent theory for the formation and late evolution of planetary systems.
Although the predominant focus of exoplanetary science has been MS systems, the last decade has
seen a surge in interest for giant branch (GB) stars and white dwarfs (WD) that host planetary systems.

\subsection{Importance of post-MS planetary systems}

The discovery of over 50 exoplanets and nearly 100 substellar companions to stars which have left the main sequence
\citep[see][and references therein]{wanetal2014} help motivate studies which attempt to link
the past and future evolution of planetary systems.  Detections of these
companions have become more robust with the combined efforts of Doppler radial velocity observations
and transit-based photometry \citep{liletal2014,cicetal2014,ortetal2014}.  Some GB stars contain multiple
planets \citep[e.g.][]{nieetal2014}, and/or enhanced abundances of lithium, which could represent
a key tracer of recent engulfment \citep[][]{adaetal2012,nowetal2013,adaetal2014}.  Observations 
of dusty debris discs around subgiant stars \citep{bonetal2013,bonetal2014} suggest that collections of 
material, such as the asteroid belt, can survive the entire MS lifetime of a star.

These observations are bolstered by strong supporting evidence for asteroidal material in 
WD systems.  Although only a few dusty debris discs are known around subgiant stars, over
35 of such discs have been observed orbiting WDs \citep{zucbec1987,becetal2005,kiletal2005,
reaetal2005,faretal2009,beretal2014,rocetal2014}.  Sometimes accompanying the dust
are gaseous components \citep{ganetal2006,ganetal2007,ganetal2008,gansicke2011,dufetal2012,
faretal2012,meletal2012} with both components overlapping radially 
\citep{brietal2009,meletal2010,brietal2012,wiletal2014,wiletal2015}.
The radial extent of these intriguing discs typically does not exceed one Solar radius,
which is close to the tidal disruption radius for an asteroid.
The environment within this compact region of space is dynamic: the gaseous
components of the discs provide kinematic information, and demonstrate sometimes
extreme variations with every new observation \citep{ganetal2008,wiletal2014,manetal2015}.
Sharp decreases in luminosity in dusty disc systems also indicate quick drastic
changes in the morphology of the circumstellar debris \citep{xujur2014}.

WDs accrete the disc material, which appear as atmospheric metal abundances, 
or ``pollution''.  Any metals heavier than helium
sink out of sight on time scales of days to Myrs (depending on the depth 
of the convection zone), which are in all cases much shorter than the WD cooling 
age, i.e the time since the star became a WD
\citep[see Fig. 1 of][]{wyaetal2014}.  Therefore, The detection of photospheric 
metals implies recent, or ongoing accretion from a circumstellar debris reservoir.
Further, between 25 to 50 per cent of all WDs 
contain signatures of metal pollution \citep{zucetal2003,zucetal2010,koeetal2014},
and at widely varying (over 5 Gyr) cooling ages.

\subsection{The link between GB and WD planetary systems}

These ubiquitously active environments close to WDs must be linked with planetary
system evolution during the GB phases of evolution.
The interstellar
medium is too diffuse to account for amount of metals seen, and is predominately 
composed of hydrogen.  Atmospheric abundances of WDs which are hydrogen-poor and 
metal-rich (known as DBZ WDs) cannot then arise from the interstellar medium
\citep{aanetal1993,frietal2004,jura2006,kilred2007,faretal2010,baretal2014}.

The currently-favoured explanation for the origin of atmospheric pollution and debris discs 
are asteroids which are dynamically perturbed close to the WD and then tidally disrupted
there.  By asteroids we refer to any objects whose radius is approximately between
0.1km-1000km lying within thousands of au of their parent stars; we denote pebbles and 
planets as objects with radii between 1mm-1m and larger than 1000km, respectively. 
Evidence suggests 
that other potential explanations are unlikely.  
The accretion and
disruption of planets around WDs likely does occur, but too infrequently to explain the large fraction
of all WDs currently-observed to be accreting \citep{veretal2013a,musetal2014,vergae2015}.
Exo-Oort cloud comets (with semimajor axes of tens of thousands or hundreds of thousands of au) are unlikely debris progenitors on both chemical \citep{zucetal2007} and dynamical
grounds \citep{veretal2014a}, although in isolated cases they may produce detectable accretion
\citep{stoetal2015}.

Asteroids provide a readily-available reservoir of material with a range of metals 
diverse enough to help explain the elemental medley observed in atmospheric pollution 
\citep{dufetal2012,gaeetal2012,xuetal2014}.  Further, the water retention level in asteroids during
the GB phases is uncertain \citep{jurxu2010,jurxu2012}, but might be high enough to explain
WDs with water-rich atmospheres \citep{faretal2013,radetal2015}.

Dynamically, asteroids can be scattered onto or close to the WD through gravitational interactions
with surviving planets.  Subsequently, the asteroids which avoid direct collisions might tidally
disrupt \citep{graetal1990,jura2003,debetal2012,beasok2013,veretal2014b}, creating discs and 
accreting onto the WD \citep{rafikov2011a,rafikov2011b,rafgar2012,wyaetal2014}.
Some investigations which modelled asteroid perturbations during and beyond
GB mass loss have explored different regions of the available phase space, and included one 
planet in their simulations.  These studies encompass \cite{bonetal2011}, who considered asteroids 
in an exo-Kuiper belt (at 30 au from the star), and \cite{debetal2012} and \cite{freetal2014}, who 
modelled a belt more akin to the asteroid belt (at about a few au from the star).

\subsection{Previous most relevant work}

None of those three studies included effects from radiation or stellar wind drag in their numerical simulations 
during GB evolution.  However, \cite{veretal2014c} demonstrated that nearly all asteroids with radii between
100m and 10km within about 7 au of their parent star on the MS will later be destroyed because of rotational
fission -- due to the GB star's luminosity -- from the YORP effect.  Consequently, an exo-belt 
similar to the asteroid belt simply will not survive the GB stellar phases intact,
and this phenomenon should be taken into account when modelling asteroid evolution during 
the WD phases.  

Yet, YORP-induced fission is just one consequence of the violent dynamical environment
of GB stars.  \cite{bonwya2010} described several others in the context of their influence
on debris discs orbiting stars which turn off of the main sequence.  The authors considered the effects
from Poynting-Robertson drag, and forces they denoted as radiation pressure, stellar wind pressure,
and stellar wind drag.  Here, we isolate these forces and
derive how they change the orbits of small bodies in a general fashion, without making any 
assumptions about size distributions 
other than that their diameters are large compared to the wavelength of the incident radiation.  
We determine, for example,
the eccentricity and inclination evolution of orbiting bodies; \cite{bonwya2010}
assumed that the bodies were on circular coplanar orbits.

In addition to \cite{bonwya2010}, \cite{donetal2010} also considered physical
effects arising from GB stars. \cite{donetal2010} focused on 
stellar wind drag and entrainment of small particles along the GB phases, and explicitly 
included drag in their equations of motion (Stokes regime only), but did not 
present those equations in orbital elements.  They neglected 
Poynting-Robertson drag, YORP and Yarkovsky\footnote{\cite{donetal2010} do not make a distinction
between the Yarkovsky and YORP effects.} effects during AGB evolution, 
but estimated that those effects would lead to significant orbital evolution - a notion 
that we quantify in detail here.

\begin{figure}
\centerline{
\psfig{figure=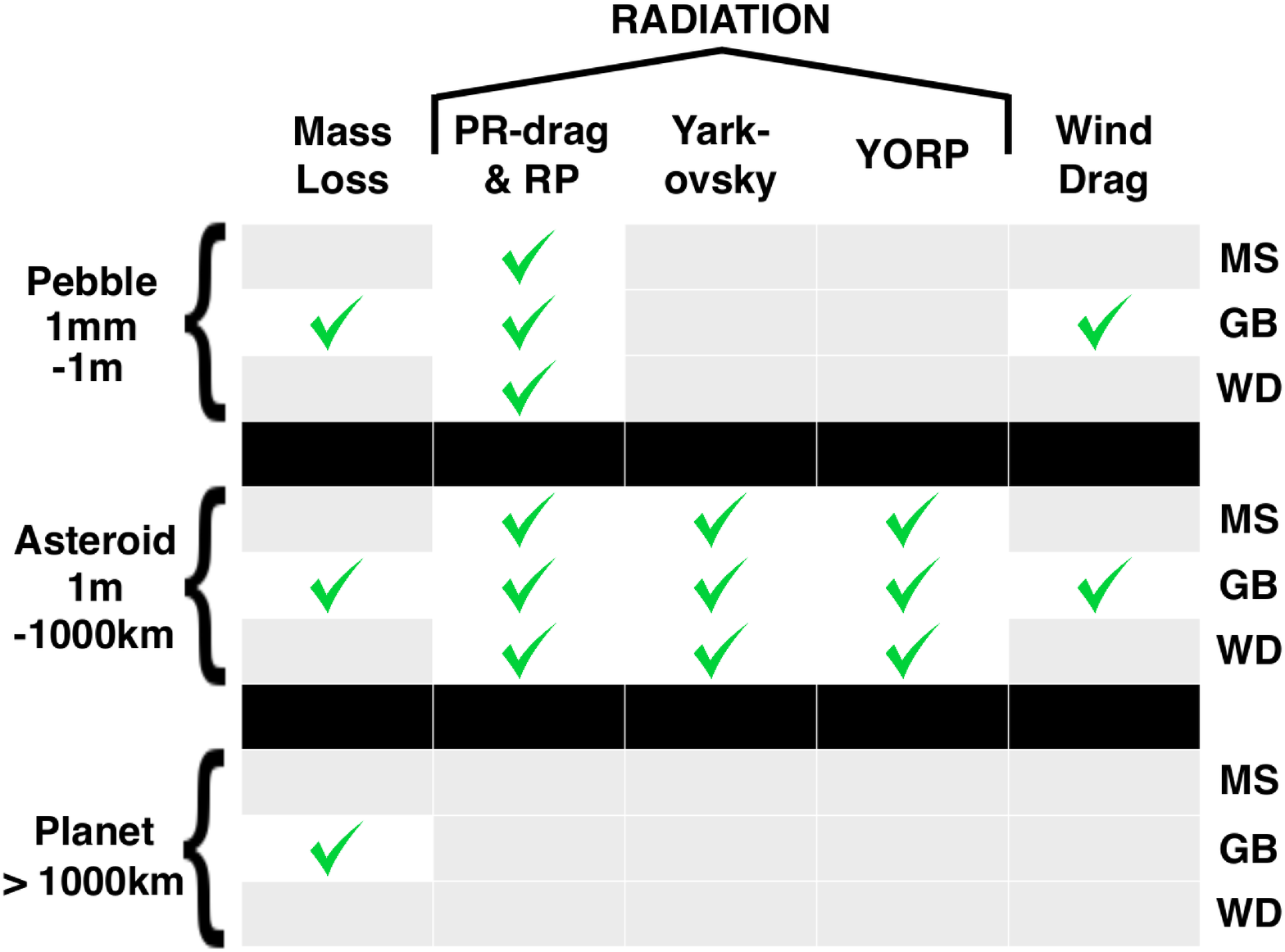,height=6.9cm} 
}
\caption{Effects which are necessary to consider for modeling.
The designations MS, GB and WD refer to the main
sequence, giant branch and white dwarf phases of 
stellar evolution.  The radiative forces include 
those from ``PR-drag and RP'' 
(Poynting-Robertson drag and radiation pressure;
see Section 2.5), the Yarkovsky effect
(see Section 2.6), and the YORP effect
(see Section 2.2).  Properties of stellar winds
are introduced in Section 3.3.
}
\label{Sumchart}
\end{figure}

Before proceeding with the body of the paper, we present a summary chart
(Fig. \ref{Sumchart}) of the physical effects which should be considered when 
modeling pebbles, asteroids or planets along different phases of stellar evolution.
Boxes without checkmarks indicate forces that are negligible along that particular
phase.  

This chart simultaneously showcases our conclusions and reaffirms some results
by both \cite{bonwya2010} and \cite{donetal2010}.  For example, both studies
find that the distribution of pebbles in WD systems will be crucially determined by
wind drag along the GB phase.  Multiple checkmarks in a single row of Fig.  \ref{Sumchart}
indicate either that multiple forces must be considered together for that stellar
phase, or that one may dominate the motion, depending on the physical and orbital 
variables chosen.  Additional forces not included in the chart, such as 
collisions \citep{bonwya2010} or those from additional planets \citep{donetal2010}, can
alter the dominant force in a given system during a given phase.

\subsection{Outline for this article}

Here we determine how GB stars change the orbits of 
bodies in an all-inclusive fashion under the guise of the
perturbed two-body problem.  We perform direct comparisons of the effects
due to gravity, radiation and drag.  We then derive equations of motion
in orbital elements through the equation-generation procedure 
outlined in \cite{vereva2013a}.  All of the variables and parameters
used are delineated in Tables 1-2 for easy reference.
Henceforth, we denote the
bodies as {\it targets} to help avoid any bias
towards pebbles, asteroids or planets.
Let $m$ represent the target's fixed mass, $\vec{v}$ the velocity of the target with respect to the star, 
and $\vec{r}$ the distance between the centre of the target and the centre of the 
star\footnote{This distance is between the centres
of the bodies and not their surfaces because we will assume Gauss' law to calculate the incident 
radiation on the asteroid.  Therefore, as long as the target remains outside of the 
star's surface and the stellar emission does not significantly vary over the star's surface,
our equations will hold.  Further, the momentum transport equations contain the equivalent
cross section of the asteroid, which is assumed to be located where the asteroid has its
largest diameter. For a spherical asteroid the cross sectional plane contains the asteroid's centre.}.  
The star has a mass of $M$ which does change with time.
The target then evolves according to

\begin{eqnarray}
\frac{d\vec v}{dt}
&=&
-
\frac{G\left(M(t=0) + m\right) \vec{r}}{r^3}
\nonumber
\\
&+&
\left(\frac{d\vec{v}}{dt}\right)_{\rm ml}
+
\left(\frac{d\vec{v}}{dt}\right)_{\rm ra}
+
\left(\frac{d\vec{v}}{dt}\right)_{\rm dr}
\label{eqofm}
\end{eqnarray}

\noindent{}where the three rightmost terms of equation (\ref{eqofm}) refer to the contributions from
stellar mass loss, radiation, and wind drag.  Cartesian formulae for wind drag and mass loss
are well-established; radiative effects provide more subtleties.  Complete sets of orbital
element formulae have been previously described only for mass loss, to our knowledge.

In Section 2, we derive the nontrivial expression for $(d\vec{v}/dt)_{\rm ra}$ in terms of Cartesian
positions and velocities by developing a framework in which to consistently treat
Poynting-Robertson drag, radiation pressure, the Yarkovsky effect and YORP.
Previous investigations almost exclusively focused on the Solar system.
Our treatment here removes the biases of constant luminosity and asteroid belt
location which are inherent in those studies.

Section 3 contains our comparison of the magnitudes and relevant
size regimes of all of the terms in equation (\ref{eqofm}).   These 
comparisons of instantaneous accelerations help quantify the relative
importance of the different forces at a given time, but not over
long times (secular timescales).

In Section 4 we derive the resulting (unaveraged) equations of motion in orbital 
elements for all of these effects, when viable. We then derive the averaged
equations of motion, which reveal secular aspects of the motion.  These equations 
provide a way for us to either compute specific trajectories or place bounds on the
motion given our framework in Section 2. We summarize our findings in Section 5.

\begin{table*}
 \centering
 \begin{minipage}{180mm}
  \caption{Roman variables and parameters used in this paper.}
  \begin{tabular}{@{}llc@{}}
  \hline
   Variable & Explanation & Equation \# \\
 \hline
 $a$ & Target's semimajor axis & \\[2pt]
 $A$ & Target's cross-sectional area (assumed to be $=\pi R^2$ here) & \\[2pt]
 $B$ & Auxiliary piecewise variable & \ref{Resplit} \\[2pt]
 $c$ & Speed of light & \\[2pt]
 $C_1-C_{11}$ & Auxiliary variables & \ref{auxCs}, \ref{C10C11} \\[2pt]
 $\mathcal{C}$ & Target's specific heat capacity & \\[2pt]
 $d$ & Target's diameter & \\[2pt]
 $\vec{D}$ & Auxiliary vectors & \ref{auxDc}-\ref{auxDomega} \\[2pt]
 $e$ & Target's eccentricity & \\[2pt]
 $f$ & Target's true anomaly & \\[2pt]
 $\mathcal{D}$ & Target's penetration depth scaling & \ref{pendep} \\[2pt]
 $G$ & Gravitational constant &  \\[2pt]
 $\vec{h} $ & Target's specific orbital angular momentum ($= \vec{r} \times \vec{v}$) &  \\[2pt]
 $i$ & Target's inclination with respect to the star's equator & \\[2pt]
 $\mathbb{I}$ & 3x3 unit matrix & \\[2pt]
 $k$ & Constant between 0 and 1/4 &  \ref{therm2} \\[2pt]
 $K$ & Target's thermal conductivity & \\[2pt]
 $\mathcal{K}$ & Constant equal to 0.165 & \\[2pt]
 $L$ & Star's luminosity & \\[2pt]
 $M$ & Star's mass  & \\[2pt]
 $m$ & Target's mass & \\[2pt]
 $m_{\rm H}$ & Mass of Hydrogen atom ($\approx 1.66 \times 10^{-27}$ kg) & \\[2pt]
 $n$ & Target's mean motion  & \\[2pt]
 $p$ & Momentum  & \\[2pt]
 $P$ & Power  & \\[2pt]
 $Q_{\rm abs}$ & Target's absorption efficiency (dimensionless) & \\[2pt]
 $Q_{\rm ref}$ & Target's reflecting efficiency, or albedo (dimensionless) & \\[2pt]
 $Q_{\rm yar}$ & Difference of $Q_{\rm abs}$ and $Q_{\rm ref}$ (dimensionless) & \\[2pt]
 $Q_{\rm PR}$ & Target's radiative efficiency due to Poynting-Robertson drag (dimensionless) & \ref{PRQ} \\[2pt]
 $\mathbb{Q}$ & 3x3 diagonal radiation matrix & \ref{fancyQ} \\[2pt]
 $\vec{r}$ & Distance between the centre of the target and the centre of the star & \\[2pt]
 $R$ & Target's radius & \\[2pt]
 $R_{\star}$ & Star's radius & \\[2pt]
 $\mathcal{R}$ & Auxiliary variable & \ref{mathcalReq} \\[2pt]
 $\mathbb{R}$ & General 3x3 rotational orbital matrix & \ref{rotgen1}-\ref{rotgen9} \\[2pt]
 $\mathbb{R}_1\left(\vec{s}\right)$ & 3x3 Rotational matrix \#1 for target spin & \ref{R1s} \\[2pt]
 $\mathbb{R}_2\left(\vec{s}\right)$ & 3x3 Rotational matrix \#2 for target spin & \ref{R2s} \\[2pt]
 $\mathbb{R}_1\left(\vec{h}\right)$ & 3x3 Rotational matrix \#1 for orbital angular momentum & \ref{r1hmat} \\[2pt]
 $\mathbb{R}_2\left(\vec{h}\right)$ & 3x3 Rotational matrix \#2 for orbital angular momentum & \ref{r2hmat} \\[2pt]
 $\mathbb{R}_{\rm Y}\left(\vec{s},\phi\right)$ & 3x3 Rotational matrix for total diurnal Yarkovsky contribution & \ref{roths} \\[2pt]
 $\mathbb{R}_{\rm Y}\left(\vec{h},\xi\right)$ & 3x3 Rotational matrix for total seasonal Yarkovsky contribution & \ref{rothxi} \\[2pt]
 ${\rm Re}$ & Reynolds number of stellar wind at the target's location & \ref{ReNum} \\[2pt]
 $\vec{s}$ & Target's spin axis (with angular speed = $|\vec{s}|$) & \ref{sdef} \\[2pt]
 $s_{\rm crit}$ & Angular speed beyond which bodies in the Solar system with $0.1$ km $\lesssim R \lesssim 10.0$ break up & \ref{scdef} \\[2pt]
 $T$ & Temperature  & \ref{TempEq} \\[2pt]
 $U_{\odot}$ & A fiducial Solar force ($= 10^{17}$ kg m/s$^2$) & \ref{sdef} \\[2pt]
 $\vec{v}$ & Target's velocity with respect to the star & \\[2pt]
 $\vec{v}_{\rm g}$ & Gas velocity (where the gas is the stellar wind) & \\[2pt]
 $v_{\rm s}$ & Local sound speed of gas (stellar wind) & \\[2pt]
 $V_{\rm rot}$ & Star's rotational velocity at its equator & \\[2pt]
 $W$ & Coefficient of the target's physical asymmetry (dimensionless) & \ref{sdef} \\[2pt]
 $\mathbb{Y}$ & 3x3 diagonal Yarkovsky matrix & \ref{fancyY} \\[2pt]
 $Z$ & Stellar latitude & \ref{gasspeed} \\[2pt]
\hline
\end{tabular}
\end{minipage}
\end{table*}

\begin{table*}
 \centering
 \begin{minipage}{180mm}
  \caption{Greek variables and parameters used in this paper.}
  \begin{tabular}{@{}llc@{}}
  \hline
   Variable & Explanation & Equation \# \\
 \hline
 $\beta$ & Ratio of target's acceleration from radiation to that from gravity & \ref{betaeq} \\[2pt]
 $\gamma$ & Ratio of target's acceleration from radiation to that from mass loss & \ref{gammaeq} \\[2pt]
 $\delta$ & Ratio of target's acceleration from radiation to that from wind drag & \ref{deltaeqe}-\ref{deltaeqs} \\[2pt]
 $\epsilon$ & Target's emissivity & \\[2pt]
 $\zeta$ & Mean free path length of gas (stellar wind) & \\[2pt]
 $\theta$ & Collision cross-section of gas molecules & \\[2pt]
 $\vec{\iota}$ & Relativistically-corrected direction of incoming radiation & \ref{PRBurns2}\\[2pt]
 $\kappa$ & Auxiliary variable & \ref{kappaeq}\\[2pt]
 $\bar{\lambda}$ & The peak wavelength of absorbed radiation & \\[2pt]
 $\mu$ & Mean molecular weight & \\[2pt]
 $\xi$ & Seasonal thermal lag angle in the orbital plane  & \\[2pt]
 $\Pi$ & Target's orbital period ($= 2\pi/n$) & \\[2pt]
 $\rho$ & Target's mass density & \\[2pt]
 $\rho_{\rm g}$ & Mass density of gas (stellar wind) & \\[2pt]
 $\sigma$ & The Stefan-Boltzmann constant & \\[2pt]
 $\Sigma$ & Target's rotational spin period ($= 2\pi/|\vec{s}|$) & \\[2pt]
 $\Upsilon_1-\Upsilon_9$ & Auxiliary variables & \ref{domegadtavgYark} \\[2pt]
 $\phi$ & Dirurnal thermal lag angle along the target's equator & \\[2pt]
 $\Psi$ & Mass loss index [$= (dM/dt)/(nM)$ ] & \ref{Psidmdt} \\[2pt]
 $\omega$ & Target's argument of pericentre & \\[2pt]
 $\Omega$ & Target's longitude of ascending node & \\
\hline
\end{tabular}
\end{minipage}
\end{table*}

\section{A framework to model radiative effects}

Over the last century, the movement of a Solar system object due to radiation pressure 
has popularly been described through Poynting-Robertson drag 
\citep{poynting1904,robertson1937}, the Yarkovsky effect \citep{radzievskii1954,peterson1976}
and the YORP (Yarkovsky-O'Keefe-Radzievskii-Paddack) effect 
\citep{radzievskii1954,okeefe1976,paddack1969}
\footnote{Yarkovsky's original paper, dating from around 1900, has apparently been lost
\citep{opik1951}.}.  Poynting-Robertson drag and the Yarkovsky effect, which are rarely
mentioned in the same context (see \citealt*{peterson1976} for an exception), both refer to 
a type of {\it orbital} recoil effect. 
YORP instead describes the change in the {\it spin} of
a body due to absorbed radiation. These notions are quantified more precisely
in the following subsections.

Although all three effects were originally defined in a Solar system context, their usage and 
names have been extended to extrasolar systems, and particularly post-main-sequence 
exosystems \citep[e.g.][]{rafikov2011a,rafikov2011b,veretal2014c}.  Their applicability
to extrasolar systems is at a fundamentally different level partly because of the difference
in the precision of available data.  Numerous studies 
\citep[e.g.][]{kry2013,rozetal2013,jacetal2014,lowetal2014,luptie2014,polishook2014,botetal2015} 
performed simulations to match outcomes to known solar system objects at a standard which will be 
unobtainable in
exosystems in the foreseeable future.  Therefore, this paper emphasizes placing bounds on
the motion and learning as much as possible from the equations of motion, rather than concerning
itself with detailed simulation suites.  But first, we must create a self-consistent model that
is free from the Solar system biases of a constant-luminosity star and asteroids concentrated
in location-specific families.

\subsection{Goal}

Our goal in Section 2 is to find $(d\vec{v}/dt)_{\rm ra}$, which can be divided into the 
following three terms

\begin{equation}
\left(\frac{d\vec{v}}{dt}\right)_{\rm ra}
\equiv
\left(\frac{d\vec{v}}{dt}\right)_{\rm abs}
+
\left(\frac{d\vec{v}}{dt}\right)_{\rm ref}
+
\left(\frac{d\vec{v}}{dt}\right)_{\rm yar}
.
\label{dvdt}
\end{equation}

\noindent{}The first term on 
the RHS of equation (\ref{dvdt}) refers to the extra acceleration on
the target due to the absorption of radiation.  The second and third
terms together give the acceleration due to the re-emission of radiation.
The second term refers to re-emission from immediate reflection, 
and the third term to re-emission from delayed reflection.  This delayed reflection
is the Yarkovsky effect, and arises from internal thermal redistribution
of the absorbed radiation.  Note that YORP is absent from equation
(\ref{dvdt}) because it provides no direct orbital acceleration to the target.

\subsection{Acceleration due to YORP}

Nevertheless, crucially, YORP
spins up or down the target.  Assume that the target rotates at an
angular speed $s$ around its spin axis, defined as 
$\vec s=\{s_x,s_y,s_z\}^T$.  Then YORP causes the angular speed to evolve
according to \citep{scheers2007}

\begin{equation}
\frac{ds}{dt} = 
        \frac{W}{2 \pi \rho R^2}
\left(
        \frac{1}{a^2 \sqrt{1 - e^2}}
\right)
\left(
        U_{\odot}\frac{L(t)}{L_{\odot}}
\right)
\label{sdef}
\end{equation}

\noindent{}where we gave the star's luminosity $L$ a time dependence for
emphasis, $0 \le W < 1$ is a constant determined by the physical
properties (roughly the asymmetry) of the target, $R$ is the 
target's radius, $\rho$ is the target's density, $a$ is the
semimajor axis of the orbit, and $e$ is the eccentricity of the orbit.
The term in the last parenthesis comes from \cite{veretal2014c} and
represents a fiducial Solar force 
($U_{\odot} = 10^{17}$ kg m/s$^2$) for application to general extrasolar
systems.

Equation (\ref{sdef}) is important because it helps determine 
if a target will undergo fission or survive.  In the latter
case, the spin evolution could affect the target's acceleration through
the Yarkovsky effect, as described in the below sections.
Solar system observations
show a sharp cutoff for the maximum spin of asteroids with radii
approximately between 0.1 km and 10 km \citep{harris1994,jacetal2014}.  
This critical spin, $s_{\rm crit}$, corresponds to a critical breakup period
$\Sigma_{\rm crit}  = 2\pi/s_{\rm crit} \approx 2.33$ hrs.
If the target is
assumed to be a strengthless rubble pile\footnote{ 
Recently, \cite{sansch2014} derived spin barriers for bodies
which harbour weak cohesive strength due to van der Waals forces between
constituent grains.  The critical breakup period is similar, but
not monotonic as a function of decreasing target radius.
}, then we can express
$s_{\rm crit}$ in terms of target density $\rho$ as

\begin{equation}
s_{\rm crit} \equiv \frac{2\pi}{\sqrt{3\pi/G\rho}} = 
7.48 \times 10^{-4} \left( \frac{\rho}{2 {\rm \ g}/{\rm cm}^3 } \right)^{\frac{1}{2}} \frac{{\rm rad}}{\rm s}
.
\label{scdef}
\end{equation}

The interplay between the YORP and Yarkovsky effects is complex.  For example,
YORP may spin up a target quickly enough to activate the Yarkovsky effect.  However,
if the target is spinning too quickly, it will lose its thermal gradient (or break up).  
We do not consider the time evolution of the spin in this work.

\subsection{Acceleration due to absorbed radiation}

The expression for the acceleration due to absorbed radiation is from equation 2 of
\cite{buretal2014}

\begin{equation}
\left(\frac{d \vec{v}}{dt}\right)_{\rm abs}
=
\frac{AL(t)Q_{\rm abs}}{4\pi m c r^2}
\vec{\iota}
\label{PRBurns1}
\end{equation}

\noindent{}where $A$ is the geometric cross-sectional area of the
illuminated surface, $Q_{\rm abs}$ represents the target's absorption efficiency
(or, equivalently, fractional amount of energy absorbed that contributes to the
momentum transfer), $c$ is the speed of light, 
and $\vec{\iota}$ is the relativistically-corrected direction of the incoming 
radiation to the target such that

\begin{equation}
\vec{\iota}
\equiv
\left(1 - \frac{\vec{v} \cdot \vec{r}}{cr} \right) \frac{\vec{r}}{r}
- \frac{\vec{v}}{c}
.
\label{PRBurns2}
\end{equation}

\noindent{}Our equation (\ref{PRBurns1}) differs from equation 2 from \cite{buretal2014}
because we consider a general extrasolar system with a time-dependent luminosity
instead of the Solar constant.  Also, \cite{buretal2014} explain that the primed quantities used in 
their equation are equivalent to the unprimed values in equation (\ref{PRBurns1}) to low order in
$v/c$; this difference in quantities has previously been identified \citep{kleetal2014}.

\subsection{Acceleration due to immediately-reflected radiation}

The acceleration on the target due to immediately-reflected radiation
is expressed in a similar form

\begin{equation}
\left(\frac{d \vec{v}}{dt}\right)_{\rm ref}
=
\frac{AL(t)Q_{\rm ref}}{4\pi m c r^2}
\vec{\iota}
\label{dirsca}
\end{equation}

\noindent{}where $Q_{\rm ref}$ is the target's reflecting efficiency,
or, the fractional amount of energy immediately reemitted after
absorption, or simply the albedo. In order to avoid having to account for 
complex scattering properties of tiny particles, 
we assume that our targets are at least 100 times the wavelength of radiation (corresponding 
to the approximate limit between geometric optics and Mie scattering regime).

\subsection{Acceleration due to Poynting-Robertson drag and ``radiation pressure''}

\cite{buretal1979} observed that the terms {\it Poynting-Robertson drag} and
{\it radiation pressure} are ambiguous.
Despite how the pressure created by radiation is responsible for {\it all} of the terms
on the RHS of equation (\ref{dvdt}), plus YORP, the term {\it radiation pressure} has 
typically come to mean just the radial component of the sum of $\left(\frac{d \vec{v}}{dt}\right)_{\rm abs}$ 
and $\left(\frac{d \vec{v}}{dt}\right)_{\rm ref}$.  Similarly, the term {\it Poynting-Robertson
drag} has typically come to mean just the tangential component of the sum of 
$\left(\frac{d \vec{v}}{dt}\right)_{\rm abs}$ and $\left(\frac{d \vec{v}}{dt}\right)_{\rm ref}$.
These components are isolated and discussed in \cite{buretal1979},
\cite{mignard1984} and \cite{hamilton1993}.

Also identified in previous studies is a $Q$ coefficient that
is associated with Poynting-Robertson drag, and is already averaged over
the stellar spectrum (see e.g. equations 4 and 19 of \citealt*{buretal1979}).  This coefficient
is related to our formalism through:

\begin{equation}
Q_{\rm PR}  = Q_{\rm abs} + Q_{\rm ref}
\label{PRQ}
.
\end{equation}

\noindent{}One must keep in mind that the values of $Q$ are averaged over the stellar
spectrum because the stellar spectrum changes with stellar evolutionary phase.  Such
changes will not cause our conclusions to deviate significantly.

\subsection{Acceleration due to the Yarkovsky effect}

\begin{figure*}
\centerline{
\psfig{figure=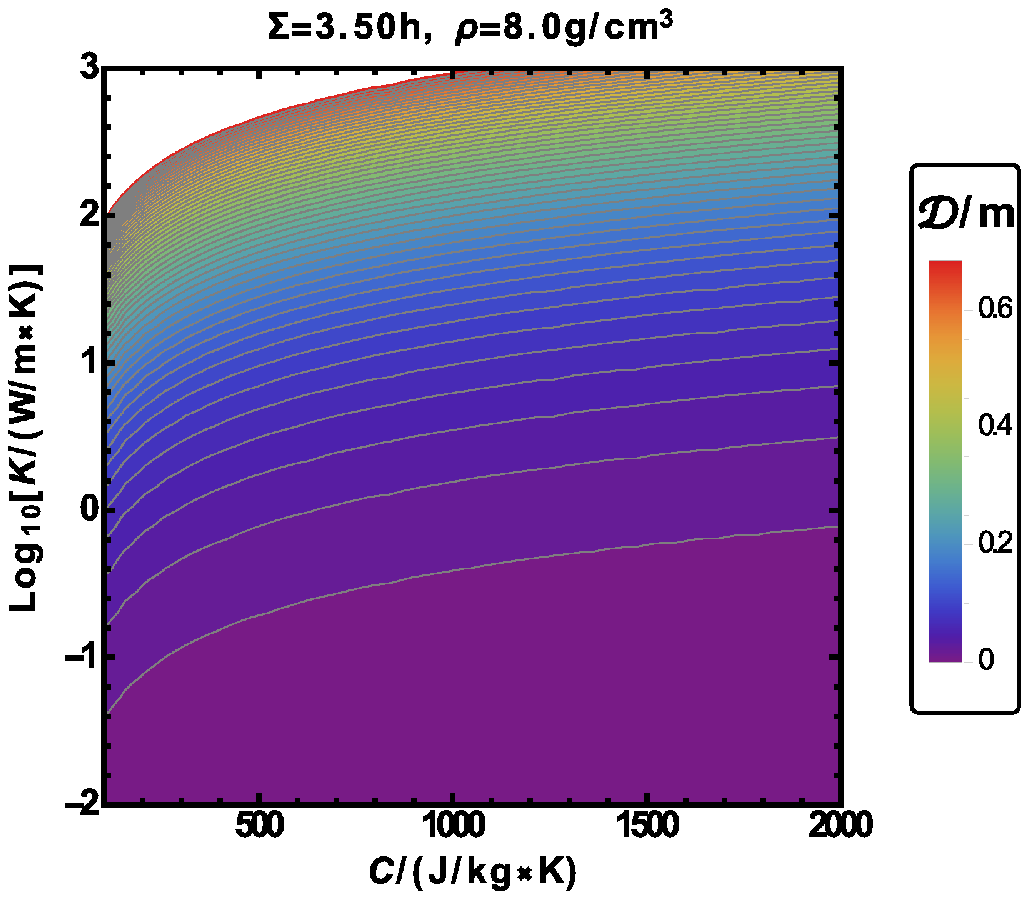,height=6.5cm} 
\ \ \ \ 
\psfig{figure=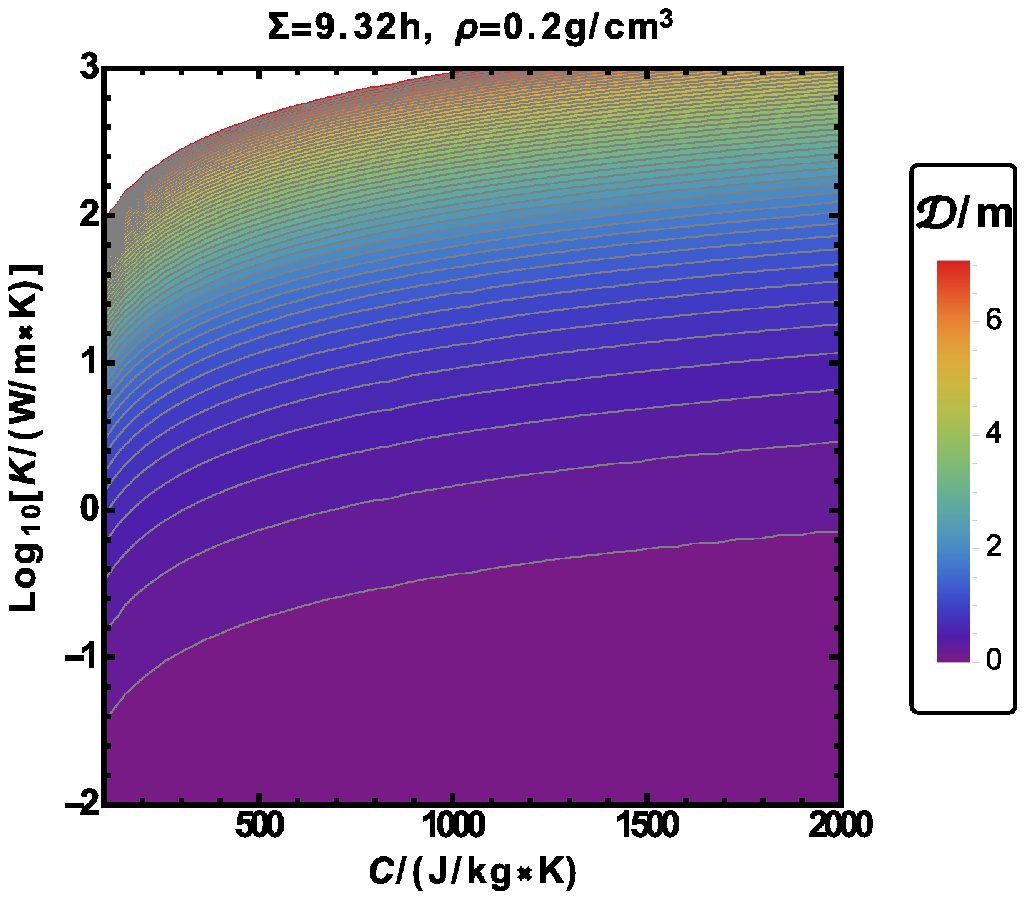,height=6.5cm} 
}
\caption{Minimum diameter of target at which the Yarkovsky effect ``turns on''.
This critical diameter is taken to equal the penetration depth scaling $\mathcal{D}$,
which is a function of the thermal conductivity $K$ and thermal capacity $\mathcal{C}$.
The left panel describes a high density 
$\rho$ target with a fast spin period, equal to $1.5 \Sigma_{\rm crit}$.
The right panel instead describes a low density slowly spinning target with 
$\Sigma = 4.0 \Sigma_{\rm crit}$.  The plots 
bound realistic values of $\mathcal{D}$ to between about 1 cm and 10 m.
}
\label{pendepfig}
\end{figure*}

Now we come to describing the last piece of equation (\ref{dvdt}), the
perturbative acceleration due to the Yarkovsky effect.  Recall that the
Yarkovsky effect is due to reemission from delayed reflection.  The
delay comes from the redistribution of thermal energy within the
target.  Therefore, if the target is too small, or spins very rapidly, no 
significant temperature differences will occur, and the Yarkovsky effect ``turns off''.

\subsubsection{Penetration depth}

The minimum size of the target at which the effect becomes negligible is approximately the
penetration depth scale of the radiation.  This depth scale can be derived from a 
heat conduction model.  Here we adopt the simplified 1D model of
\cite{broz2006}.  This model provides the time lag between insolation
and re-emission in realistic materials.  The model also assumes that
the target rotates at a constant angular speed, and therefore strictly
equation (\ref{sdef}) cannot be used in conjunction.  Realistically, 
however, YORP spin changes occur relatively slowly compared to thermal
wave properties.  Therefore, the YORP effect may be included in the model 
if $\vec{s}$ is considered to be a function of time.

This penetration depth scaling, $\mathcal{D}$, is

\begin{equation}
\mathcal{D} = \sqrt{ \frac{K\Sigma}{\pi \rho \mathcal{C}} }
\label{pendep}
\end{equation}

\noindent{}where $K$ is the thermal conductivity and $\mathcal{C}$
is the specific thermal capacity (or specific heat capacity).
Hence, the Yarkovsky effect will ``turn on'' if the target diameter
is larger than\footnote{In principle the thermal wave
can heat the target by reaching down to its core. In this case, the
sufficient condition for the Yarkovsky effect to ``turn on'' is
if the target radius is larger than $\mathcal{D}$.} $\mathcal{D}$. As this scaling 
significantly depends on the target's physical properties, we provide contour plots in
Fig. \ref{pendepfig}.  The contour plots cover the entire range of
possible $\mathcal{D}$ values for both a high density, quickly
spinning (left panel) target and a low density, slowly spinning
target (right panel).  Note that 8 g/cm$^3$ is slightly higher than 
the density of iron and 0.2 g/cm$^3$ is slightly less dense than 
cork.  Target thermal conductivities are confined to
$10^{-2}-10^3$ W/(m$\times$K), which include all substances from air
to silver.  Target specific heat capacities are confined to about 
$100$ J/(kg$\times$K) for heavy metals to about $2000$ J/(kg$\times$K)
for water.  The result, in all cases, is
1 cm $\lesssim \mathcal{D} \lesssim$ 10 m, conforming well to Figs. 25-26
of \cite{broz2006}.  Therefore, we do not expect
Yarkovsky to be active in sub-cm-sized targets, and we 
expect Yarkovsky to be always active in moderately fast spinning targets larger than 10 m.

\begin{figure}
\centerline{
\psfig{figure=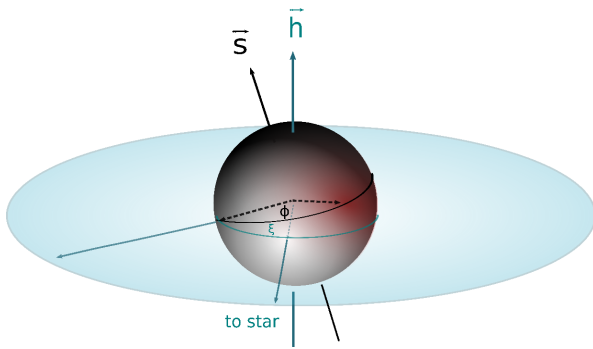,height=4.5cm} 
}
\caption{Relevant angles and vectors for the 
target.  The spin axis is $\vec{s}$, the specific
angular momentum axis is $\vec{h}$, $\phi$ is
the thermal lag angle along the target's equator,
and {$\xi$ is the thermal lag angle in the orbital plane.  The 
white spot corresponds to the direction of the star (subsolar point), and 
the red spot to the maximum of the thermal wave 
(the direction opposite to the Yarkovsky drift).}
}
\label{geomchart1}
\end{figure}

\subsubsection{Yarkovsky direction}

The geometry of the thermal redistribution determines the direction
of the Yarkovsky acceleration.  This redistribution is dependent on
the orientations of both the spin axis $\vec{s}$ and the specific
orbital angular momentum axis $\vec{h} = \vec{r} \times \vec{v}$.
Let $\phi$ and $\xi$ represent lag angles of the thermal wave.
The variable $\phi$ represents the thermal lag angle along the target's equator, i.e.
how far the heat wave is dragged along from the subsolar point due
to the target's rotation. The variable $\xi$ represents the thermal lag 
angle in the orbital plane measured from the subsolar point. See
Fig. \ref{geomchart1} for a visual representation of these angles.

Solar system studies traditionally denoted the contribution from the spin axis
and $\phi$ as ``diurnal'' and the contribution from the 
specific angular momentum axis and $\xi$ as ``seasonal''.
These characterizations can be traced historically to the
thermal imbalance between the morning and afternoon hemispheres
(for diurnal) or the winter and summer hemispheres (for seasonal)
of a Solar system asteroid.  Mathematically, the contributions
are expressed via rotation matricies through
an angle (here $\phi$ and $\xi$) about a given
axis (here $\vec{s}$ and $\vec{h}$) as

\begin{equation}
\left( \frac{d\vec{v}}{dt} \right)_{\rm yar}^{\rm diurnal}
=
\left| \left( \frac{d\vec{v}}{dt} \right)_{\rm yar}  \right|
\mathbb{R}_{\rm Y}(\vec{s}, \phi) \vec{\iota} 
,
\end{equation}

\begin{equation}
\left( \frac{d\vec{v}}{dt} \right)_{\rm yar}^{\rm seasonal}
=
\left| \left( \frac{d\vec{v}}{dt} \right)_{\rm yar}  \right|
\mathbb{R}_{\rm Y}(\vec{h}, \xi) \vec{\iota} 
,
\end{equation}

\begin{equation}
\left( \frac{d\vec{v}}{dt} \right)_{\rm yar}
=
\left| \left( \frac{d\vec{v}}{dt} \right)_{\rm yar}  \right|
\mathbb{R}_{\rm Y}(\vec{h}, \xi)
\mathbb{R}_{\rm Y}(\vec{s}, \phi) 
\vec{\iota} 
,
\label{yardir}
\end{equation}

\noindent{}where the $\mathbb{R}$ are the following general rotational matrices

\begin{equation}
\mathbb{R}_{\rm Y}(\vec{s}, \phi) \equiv \cos{\phi}\mathbb{I} + \sin{\phi}\mathbb{R}_1(\vec{s}) 
+ \left(1 - \cos{\phi} \right)\mathbb{R}_2(\vec{s}), 
\label{roths}
\end{equation}

\begin{equation}
\mathbb{R}_{\rm Y}(\vec{h}, \xi) \equiv \cos{\xi}\mathbb{I} - \sin{\xi}\mathbb{R}_1(\vec{h}) 
+ \left(1 - \cos{\xi} \right)\mathbb{R}_2(\vec{h})
\label{rothxi}
\end{equation}

\noindent{}with

\begin{eqnarray}
\mathbb{R}_1(\vec s) \equiv
\frac{1}{\sqrt{s_{x}^2+s_{y}^2+s_{z}^2}}
\left( \begin{array}{ccc} 
            0 & -s_z& s_y \\
            s_z & 0 & -s_x \\
            -s_y & s_x & 0 \\
            \end{array}
            \right),
\label{R1s}
\end{eqnarray}

\begin{eqnarray}
\mathbb{R}_2(\vec s) \equiv 
\frac{1}{s_{x}^2+s_{y}^2+s_{z}^2}
\left( \begin{array}{ccc} 
            s_x^2 & s_x s_y& s_x s_z \\
            s_x s_y & s_y^2 & s_y s_z \\
            s_x s_z & s_y s_z & s_z^2 \\
            \end{array}
            \right), 
\label{R2s}
\end{eqnarray}

\begin{eqnarray}
\mathbb{R}_1(\vec h) \equiv
\frac{1}{\sqrt{h_{x}^2+h_{y}^2+h_{z}^2}}
\left( \begin{array}{ccc} 
            0 & -h_z& h_y \\
            h_z & 0 & -h_x \\
            -h_y & h_x & 0 \\
            \end{array}
            \right),
\label{r1hmat}
\end{eqnarray}

\begin{eqnarray}
\mathbb{R}_2(\vec h) \equiv 
\frac{1}{h_{x}^2+h_{y}^2+h_{z}^2}
\left( \begin{array}{ccc} 
            h_x^2 & h_x h_y& h_x h_z \\
            h_x h_y & h_y^2 & h_y h_z \\
            h_x h_z & h_y h_z & h_z^2 \\
            \end{array}
            \right).
\label{r2hmat} 
\end{eqnarray}

In the specific 1D heat conduction model of \cite{broz2006}, the angles have the following
explicit closed forms 

\[
\tan \phi = 
\]

\[
\left(1+\frac{1}{2}\left(\frac{\sigma \epsilon}{\pi^5}\right)^{1/4} 
\left(\frac{\Sigma}{C K \rho}\right)^{1/2}\left(\frac{L(t)Q_{\rm yar}}{r^2}\right)^{3/4}\right)^{-1},
\]

\begin{equation}
\label{phiang}
\end{equation}

\[
\tan \xi = 
\]

\[
\left(1+\frac{1}{2}\left(\frac{\sigma \epsilon}{\pi^5}\right)^{1/4} 
\left(\frac{\Pi}{C K \rho}\right)^{1/2}\left(\frac{L(t)Q_{\rm yar}}{r^2}\right)^{3/4}\right)^{-1},
\]

\begin{equation}
\label{xiang}
\end{equation}

\noindent{}where $0 \le Q_{\rm yar} \le 1$ is a value satisfying $Q_{\rm abs} = Q_{\rm ref} + Q_{\rm yar}$,
$\sigma$ is the Stefan-Boltzmann constant, $\epsilon$ is the target's emissivity,
and the orbital period $\Pi$ can be expressed in terms of the target's mean motion $n$ as 
$\Pi = 2\pi/n = 2\pi a^{3/2}/\sqrt{G\left(M + m\right)}$.  The angles $\phi$ and $\xi$ can reach physically meaningful values of between $0^{\circ}$ and $45^{\circ}$; for more details on this result, and the derivation of equations (\ref{phiang}-\ref{xiang}), see \cite{broz2006}.

\subsubsection{Yarkovsky magnitude}

Having established the direction of the Yarkovsky acceleration, 
we now determine the corresponding amplitude. This
amplitude may be expressed with momentum $p$, power $P$ and temperature $T$ using
the relativistic energy momentum equation for photons with negligible
rest mass as

\begin{equation}
\left| \left( \frac{d\vec{v}}{dt} \right)_{\rm yar}  \right|
= 
\frac{1}{m} \frac{dp}{dt} = \frac{\Delta P}{mc}
=
\frac{1}{mc}
\left[P(T_1) - P(T_2) \right]
.
\label{therm}
\end{equation}

Equation (\ref{therm}) emphasizes that the Yarkovsky effect exists only
in the presence of a thermal imbalance on the target's surface.
The magnitude of the imbalance is model-dependent.  We provide insight into our
model by considering the incoming and outgoing power.  By assuming radiative balance
\footnote{After reaching its equilibrium temperature the target is able to radiate 
all of the incoming energy eventually, so that it is no longer heating up.} 
we have

\begin{eqnarray}
P_{\rm in}
&=&
\frac{L(t) A_{\rm in}Q_{\rm yar}}{4\pi r^2}
,
\\
P_{\rm out}
&=&
A_{\rm out}\epsilon\sigma T^4 
.
\end{eqnarray}

\noindent{}Equating the incoming and outgoing power gives

\begin{equation}
T=\left(\frac{A_{\rm in}}{A_{\rm out}} \frac{L(t)Q_{\rm yar}}{4\pi \epsilon \sigma r^2} \right)^{1/4} 
.
\label{TempEq}
\end{equation}

The target's equilibrium temperature $T$ depends on ratio of $A_{\rm in}/A_{\rm out}$. 
This ratio is the key to modelling the Yarkovsky effect.  Consider first
the general case of a fast rotating spherical target 
with zero obliquity. 
Because the incoming radiation is directional and the emitted radiation is omnidirectional
and symmetric with respect to the equator and the rotation axis,  
$A_{\rm in}=\pi R^2$ corresponds to the target's cross section, whereas 
$A_{\rm out}=4\pi R^2$. Alternatively, in the limit of slow rotators 
(e.g. a $1$:$1$ spin orbit resonance) the 
target shows always the same hemisphere towards the radiation source (e.g. the Moon as seen from 
the Earth). In this case, only one hemisphere is constantly heated, and without an atmosphere or 
unusually high thermal conductivity of the body, only the illuminated hemisphere will 
be able to emit the incoming energy. Hence,
$A_{\rm out}=2\pi R^2$ and $A_{\rm in}/A_{\rm out}=1/2$.  

In the limit of fast rotation, all of the target's surface elements are heated up
to the equilibrium temperature, and hence no strong thermal gradients, and no
Yarkovsky effect, is expected.  In the limit of slow rotation, because only
half of the target's surface can be used as a radiator, the equilibrium temperature
on the day side is higher by a factor of $2^{1/4}$ than the equilibrium temperature
of the entire target.  In this case, 

\begin{eqnarray}
P(T_1) - P(T_2) 
&=& 
A_1 \epsilon \sigma (2^{1/4}T)^4 - A_2 \epsilon \sigma T^4
\nonumber
\\
&=& 
\frac{1}{4} \left( \frac{L(t) A Q_{\rm yar} }{4 \pi r^2}  \right)
.
\end{eqnarray}

\noindent{}where $A_1 = A_2 = A = \pi R^2$ are the momentum-carrying
cross-sections of the asteroid's heated and non-heated hemispheres,
respectively.  However, the violent nature of the GB environment suggests
that targets with all types of spins are expected, unless they are far from the star
\citep{veretal2014c}.  Therefore, generally, the temperature difference will   
be in between the extremes of 0 and $2^{1/4}T$.  Consequently, 

\begin{equation}
\left| \left( \frac{d\vec{v}}{dt} \right)_{\rm yar}  \right|
= 
\frac{k A L(t) Q_{\rm yar} }{4 \pi m c r^2} 
\label{therm2}
\end{equation}

\noindent{}where $0 \le k \le 1/4$.  
The value of $k$ is strongly linked to the targets's rotation.  
For objects that spin very fast, i.e. $\Sigma_{\rm crit} \rightarrow 0$, $k = 0$.  
Alternatively, for $\Sigma \rightarrow \Pi$, $k = 1/4$.

In the Solar system, the albedo of an asteroid is equal to
$(1 - Q_{\rm yar})$.  Typical asteroids have albedos between 0.1
and 0.3, meaning that usually $Q_{\rm yar}$ is closer to unity
than to zero for these objects.  

\subsubsection{Final result}

We now summarize the findings in subsections 2.6.1-2.6.3 and can express the
perturbative acceleration due to the Yarkovsky effect as

\begin{equation}
\left(\frac{d \vec{v}}{dt}\right)_{\rm yar}
=
\frac{kAL(t)Q_{\rm yar}}{4\pi m c r^2}
\mathbb{Y}\vec{\iota}
\label{yarfin}
\end{equation}

\noindent{}where

\begin{equation}
\mathbb{Y} \equiv \mathbb{R}_{\rm Y}(\vec s, \phi)\mathbb{R}_{\rm Y}(\vec h, \xi)
.
\label{fancyY}
\end{equation}


\subsection{Complete expression}

\begin{figure}
\centerline{
\psfig{figure=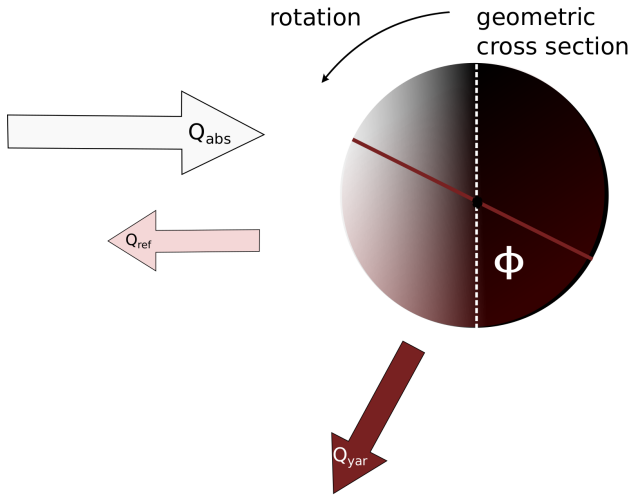,height=6.5cm} 
}
\caption{A visual summary of the relevant $Q$
values in equation (\ref{finaldvdtp}), which
represent the efficiency of absorbed radiation ($Q_{\rm abs}$),
immediately reflected radiation ($Q_{\rm ref}$), and radiation
which is reflected after a delay ($Q_{\rm yar}$).  $\phi$ is the thermal
lag angle.
}
\label{geomchart2}
\end{figure}

Finally we combine equations (\ref{PRBurns1}), (\ref{dirsca})
and (\ref{yarfin}), along with $k$ from equation (\ref{therm2}),
into the following single compact expression

\begin{equation}
\left(\frac{d\vec{v}}{dt}\right)_{\rm ra}
=
\frac{AL(t)}{4\pi m c r^2}
\bigg[  
Q_{\rm abs}\mathbb{I} + Q_{\rm ref}\mathbb{I} + kQ_{\rm yar}\mathbb{Y}
\bigg]
\vec{\iota}
,
\label{finaldvdtp}
\end{equation}

\noindent{}where the $\mathbb{I}$ and $\mathbb{Y}$ 
represent 3x3 matrices (see equation \ref{fancyY}).
These matrices reveal the relevant regimes of motion.  The first term is unchanging and
applies for targets of any size.  The third term is zeroed out for targets which are too 
small or do not spin.  If those targets are smaller than the typical wavelength of incoming 
radiation, then only the middle term is affected.  The $Q$ quantities in equation (\ref{finaldvdtp}) are all characterized visually in Figure \ref{geomchart2}.

\section{Relative importance of different accelerations}

Now that we have developed an expression that includes all of the effects
from radiation, we can estimate the relative importance of radiation
compared to mass loss, wind drag and simply unperturbed Keplerian motion
at a given time.  First, for ease of notation, define the 
term in brackets from equation (\ref{finaldvdtp}) as

\begin{equation}
\mathbb{Q}
\equiv
Q_{\rm abs}\mathbb{I} + Q_{\rm ref}\mathbb{I} + kQ_{\rm yar}\mathbb{Y}
.
\label{fancyQ}
\end{equation}

\noindent{}We now describe the physical interpretation of this model.
First the asteroid absorbs the momentum of the incoming light with 
an efficiency of $Q_{\rm abs}$.  Then part of the radiation is either
reflected immediately with an efficiency $Q_{\rm ref}$ and/or thermalized.
In the latter case the factor $kQ_{\rm yar}$ determines the relative
strength of the Yarkovsky drift.

Because for large bodies ($\lambda \ll d$), $0 \le Q_{\rm abs} \le 1$ 
and $0 \le Q_{\rm ref} \le 1$, 
each component of $\mathbb{Q}$ must lie between 0 and 2 inclusive.
Generally, $\mathbb{Q} \le 2$ corresponds to large particles like asteroids only
as smaller particles can retain scattering (and thus extinction) cross sections much 
larger than their geometric cross sections.
The Yarkovsky effect has the most importance when $Q_{\rm ref} = 0$.
In this case only a quarter of the total remitted radiative
recoil can contribute to the Yarkovsky perturbative acceleration,
as $k \le 1/4$.
However, this rudimentary analysis is deceiving, as 
we will show later with the averaged equations of motion in orbital elements.

\subsection{Radiation versus gravity}

More broadly, we can develop a sense of when radiative effects are important
by comparing the terms on the RHS of equation (\ref{eqofm}).  The first
term is the unperturbed two-body term, the second is the acceleration due to
mass loss, and the third is the acceleration due to radiation.  Define
$\beta$ as the ratio of the radiative acceleration to the total gravitational
acceleration, i.e. the ratio of the third term to the sum of the first and second.  
The sum of the first and second terms is equal to just $-G[M(t)+m]\vec{r}/r^3$,
i.e., the first term with a time dependence \citep{omarov1962,hadjidemetriou1963,deprit1983}.
Then, because
$|\vec{\iota}| \approx 1$ and $|\mathbb{Q}| \le 2$,

\begin{equation}
\beta \lessapprox
\frac{AL(t)}{2\pi G m \left(M(t) + m\right) c}
\approx
\frac{3L(t)}{8\pi G c R \rho M(t)}
,
\label{betaeq}
\end{equation}

\noindent{}an expression which is independent of the semimajor
axis $a$ and the eccentricity $e$.

\begin{figure}
\centerline{
\psfig{figure=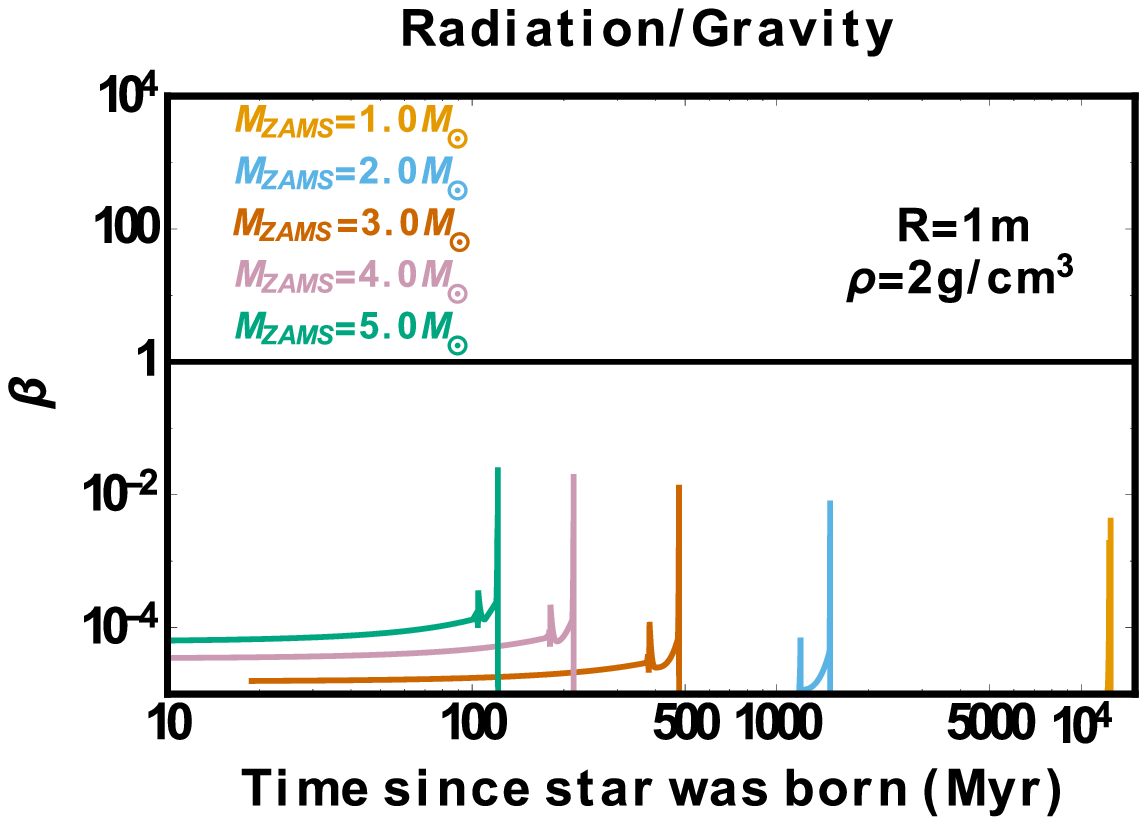,height=5.6cm,width=9.35cm} 
}
\centerline{
\psfig{figure=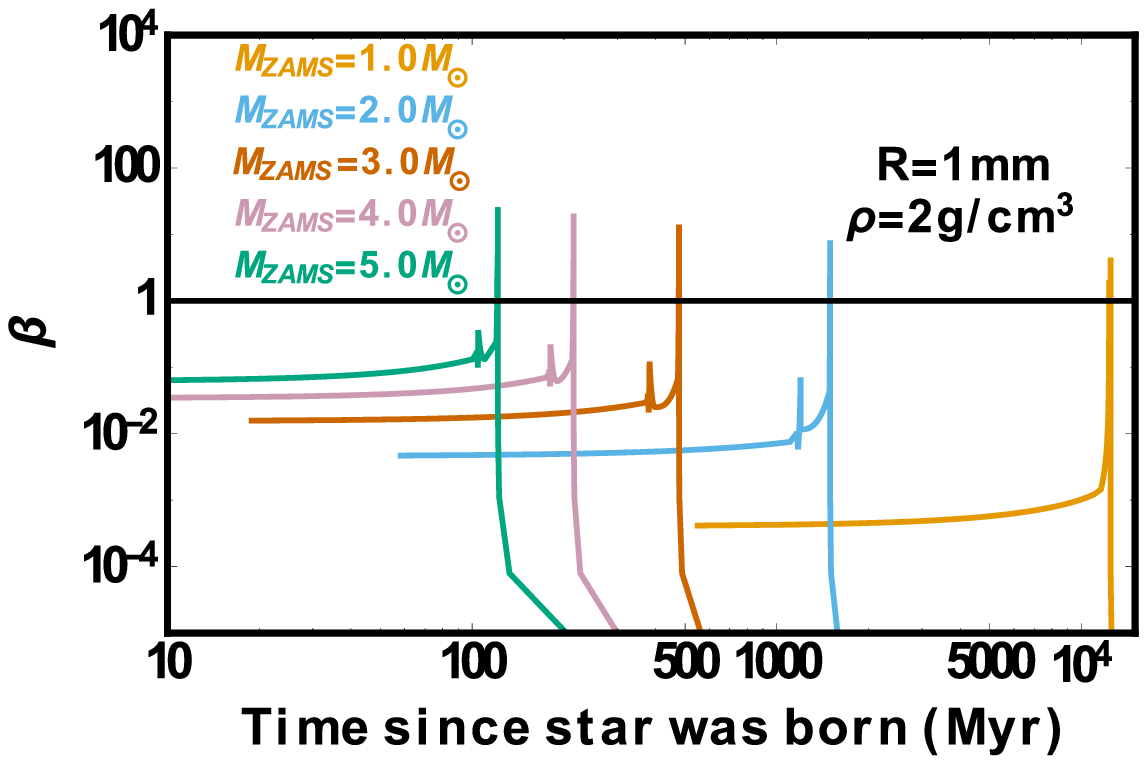,height=5.6cm} 
}
\caption{Approximate relative importance of the acceleration due to radiation 
versus that from gravity ($\beta$), as a function of stellar age, for two
different pebble-like targets with radii $R$.  The target density of 
2 g/cm$^3$ corresponds to a typical density of an icy Kuiper belt body.  Both the stellar
mass and luminosity affect $\beta$, and vary strongly with time
during the GB phases of evolution, causing the peaks in the curves.
Particles smaller than about $1$ mm will be affected more strongly
by radiation than gravity at some point in their evolution.
}
\label{betaplot}
\end{figure}

We plot $\beta$ in Figure \ref{betaplot} for targets with
radii of 1 m (top panel) and 1 mm (bottom panel), corresponding
to large and small pebbles.  The plots are a function of time, for the
entire lifetime of stars with ZAMS (zero-age main sequence)
masses of 1.0, 2.0, 3.0, 4.0 and 5.0 $M_{\odot}$.  The
variations in each curve are due to the time dependence of
both $M$ and $L$, particularly on the GB phases.
We compute stellar evolutionary tracks from the 
\textsc{sse} code \citep{huretal2000}, assuming for all stars
Solar metalicity, a Reimers mass loss coefficient of 0.5,
and a superwind on the Asymptotic Giant Branch (AGB) phase 
\citep{vaswoo1993}.

The earlier, minor peak in the curves corresponds to the tip
of the Red Giant Branch (RGB) phase, and the later peak the tip of the AGB phase.
The figure demonstrates clearly that 1mm-sized targets which
remain gravity-dominated for the entire MS and RGB phases
will suddenly become radiation-dominated during the AGB
phase. This behaviour remains true regardless of the location
of the targets.  For larger targets which are about one meter in size,
the contribution of radiation to their orbital motion
along the AGB may reach the one per cent level, potentially
enough for example to perturb the targets out of mean motion
resonance with another object - a subject for a future study
\citep[see e.g.][]{pastor2014}.

These results must be considered in the appropriate context; they
apply only for snapshots of system evolution.  In Section 4.3.4 we will
show that radiation may in fact dominate the secular evolution of
targets which are orders of magnitude larger than one meter.  The reason
is due to a potentially accumulating drift from the Yarkovsky effect,
which cannot be deduced from the ratio of instantaneous accelerations.

\subsection{Radiation versus mass loss}

The mass lost by the star, particularly during the GB phases, will
contribute to the orbital motion through a change of gravitational
potential in the system.  We can compare the orbital changes due
to radiation to those due specifically to mass loss from 
equation (\ref{eqofm}) through

\begin{equation}
\left(\frac{d\vec{v}}{dt}\right)_{\rm ml}
=
-\frac{1}{2}\frac{1}{\left(M(t) + m\right)} \frac{dM(t)}{dt}  \vec{v}
.
\label{dvdt4}
\end{equation}

\noindent{}Equation (\ref{dvdt4}) helpfully identifies the acceleration due to 
isotropic mass loss from the star in terms of positions and velocities 
\citep{omarov1962,hadjidemetriou1963,deprit1983}.  Therefore let us define $\gamma$
as the magnitude of the ratio of the acceleration due to radiation to the acceleration
due to mass loss.  We obtain

\[
\gamma \lessapprox \frac{AL(t)M(t)}{\pi m c r^2 v} \left(\frac{dM(t)}{dt}\right)^{-1}
\approx
\frac{3L}{4\pi R \rho n c r^2 v \Psi(t)}
\]

\[
\approx
\frac{3L(t)}{4\pi G M(t) R \rho c \Psi(t)}
\left[
\frac{\left(1 + e \cos{f} \right)^2}{\left(1 - e^2\right)^{3/2} \left(1 + e^2 + 2 e \cos{f} \right)^{1/2}}
\right]
,
\]

\begin{equation}
\label{gammaeq}
\end{equation}

\noindent{}where

\begin{equation}
\Psi(t) \equiv \frac{1}{nM(t)} \left| \frac{dM(t)}{dt} \right|. 
\label{Psidmdt}
\end{equation}

In equation (\ref{gammaeq}), we have introduced the true anomaly $f$ and expressed
the distance and velocity in terms of orbital elements as

\begin{eqnarray}
r &=& \frac{a\left(1-e^2\right)}{1 + e \cos{f} }
,
\\
v &=& \frac{na}{\sqrt{1-e^2}} \sqrt{1 + e^2 + 2 e \cos{f} }
.
\end{eqnarray}

\begin{figure*}
\centerline{
\psfig{figure=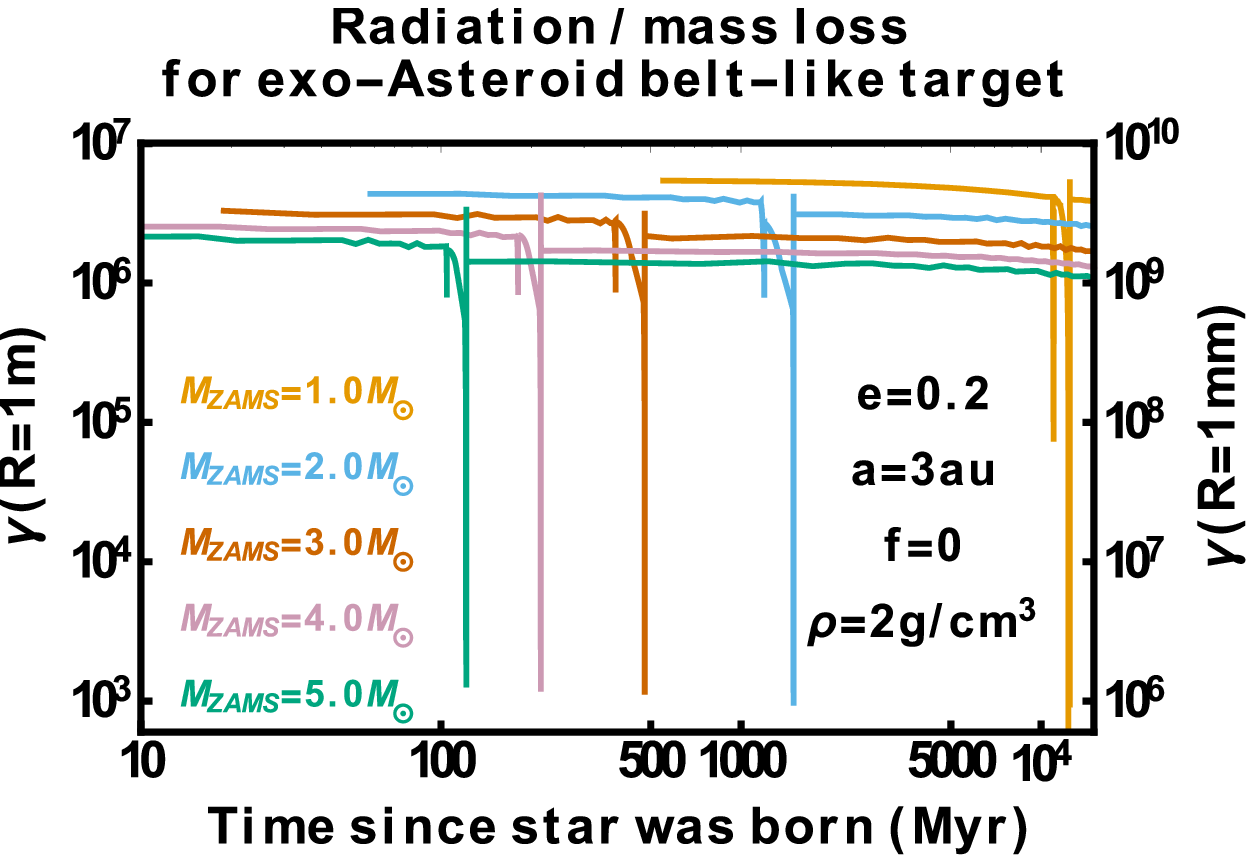,height=6cm} 
\ \ \ \ 
\psfig{figure=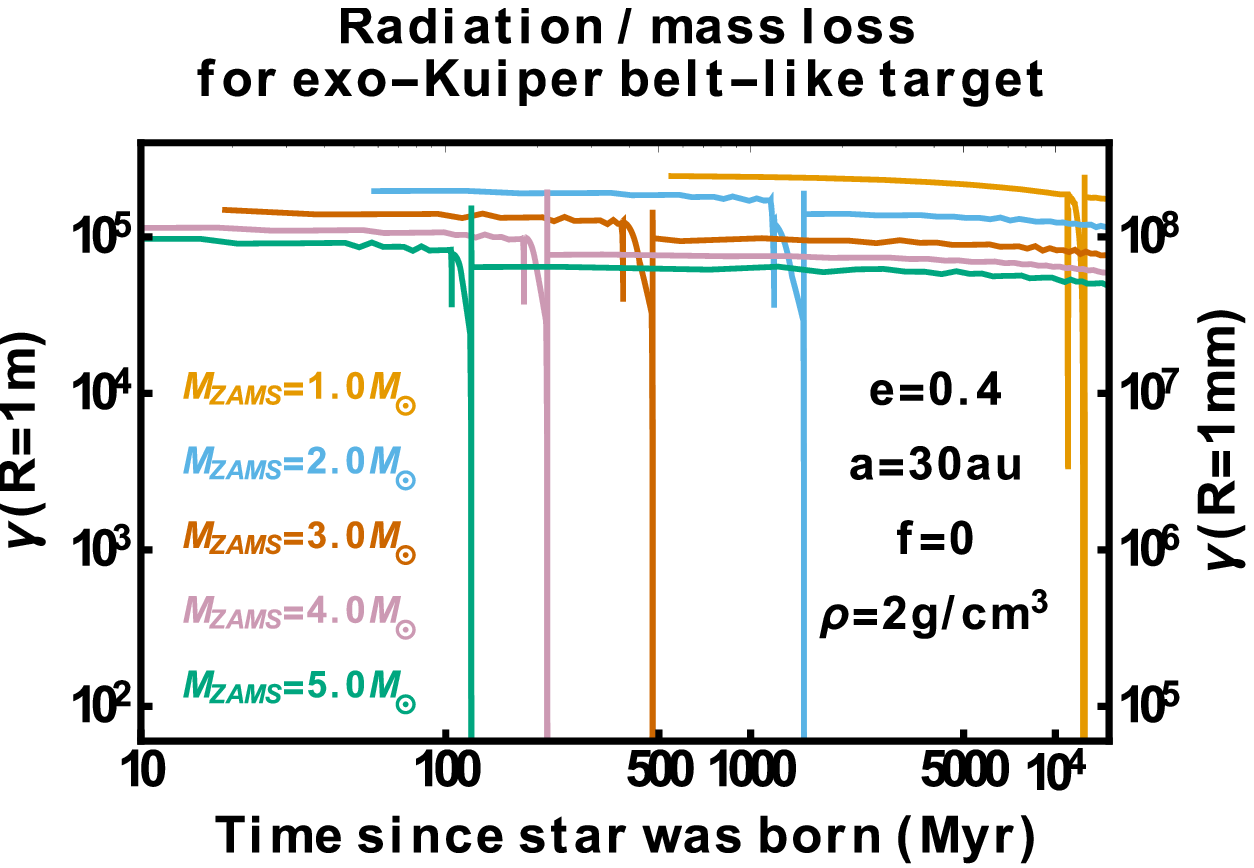,height=6cm} 
}
\caption{Relative importance of the acceleration due to radiation 
versus mass loss ($\gamma$), as a function of time, for a metre-sized
target (left $y$-axes) and a mm-sized target (right $y$-axes)
in an exo-Asteroid belt (left panel) and in an exo-Kuiper belt
(right panel).  The variables $e$, $a$, $f$ and $\rho$ refer to the
target's eccentricity, semimajor axis, true anomaly and mass density,
respectively. In all cases, orbital evolution due to radiation is
more important than that from stellar mass loss, even at the tip
of the AGB phase.
}
\label{gammaplot}
\end{figure*}

\noindent{}The dimensionless 
mass loss adiabaticity index $\Psi$ is from \cite{veretal2011}.  This index provides
a scaled ratio of the target's orbital period to mass loss timescale, and
distinguishes two distinct regimes of motion: adiabatic ($\Psi \ll 1$), when
the semimajor axis increases at a constant rate, and runaway ($\Psi \gg 1$),
where $a$, $e$, and the argument of pericentre $\omega$, increase or decrease
at nonuniform rates.  This index can be used to determine the evolution of
individual known planetary systems \citep[][]{verwya2012,musetal2013,tadetal2014},
isolate the strength of three-body interactions amidst stellar mass loss
\citep{voyetal2013}, evaluate the contribution of escaped planets to the free-floating planet population
\citep{verray2012}, and determine the relative importance of other perturbative 
effects such as Galactic tides \citep{veretal2014d}.  In equation (\ref{gammaeq}) and 
throughout the paper, we assume that the mass loss is isotropic, an excellent
approximation for targets within a few hundred au \citep{veretal2013b}.

Adopting limiting values of $f$ help us understand equation (\ref{gammaeq}).
When the target is at pericentre, the term in square brackets becomes
$(1+e)^{-1/2}(1-e)^{-3/2}$; at apocentre, this term is $(1-e)^{-1/2}(1+e)^{-3/2}$.
Therefore, for highly eccentric targets, the acceleration from radiation becomes
much more important than that from mass loss, more when the target is at
pericentre than at apocentre.  For moderate or small eccentricities, the bracketed term
may be approximated as unity, as for $e=0.5$, the term approximately equals a value between
0.8 and 2.3.  Also, unlike $\beta$, $\gamma$ is dependent on
the semimajor axis, through $\Psi(t)$.  More generally, the importance of radiation
relative to mass loss increases for (i) adiabatic motion, (ii) smaller targets,
(iii) less dense targets, and (iv) less massive stars.

We chart $\gamma$ in Figure \ref{gammaplot} for small exobody analogues of the asteroid
belt (left panel) and Kuiper belt (right panel).  The left and right $y$-axes in both
plots correspond to metre-sized and millimetre-sized targets.  We compute the mass loss index
in the same manner as in \cite{vertou2012} and assume $\Psi$ is defined in terms
of $n$ rather than $\Pi$.  This difference has a negligible impact on the resulting
curves in the figure, which span several orders of magnitude.  We consider the targets
at their pericentres; if instead we considered their apocentres, then the curves would
be shifted downward by factors of $2/3$ (left panel) and $3/7$ (right panel).

The curves suggest that for metre-sized and smaller targets within tens of au,
radiation is much more important than mass loss for orbital evolution.  Although this
ratio is most extreme on the MS, neither radiation nor mass loss contribute
much to the unperturbed Keplerian motion (given by the first term on the RHS of equation
\ref{eqofm}) during this phase.  On the AGB phase, this ratio is minimized, but
radiation is still more important than mass loss.  Therefore, when modelling 
sub-asteroidal-sized targets along the GB phases, we assert that considering only
their orbital evolution due to mass loss is insufficient.

\subsection{Radiation versus wind drag}

In the previous subsection, we focused on the gravitational evolution 
due to mass being lost from the two-body orbit.  We considered
the mass lost to be through a gaseous stellar wind, and ignored the time
lag from when the star loses mass to when the ejecta reaches
the target. Such subtleties may be unimportant for targets the size
of planets, but can be crucial for smaller targets.

\subsubsection{The different regimes of wind drag}

A solid moving through a gaseous medium will experience a drag force.
This well-studied force, in its most basic form, is proportional 
to the square of the solid's velocity.  This form has important applications
for protoplanetary discs (e.g. equation 3 of \citealt*{iwaetal2002}, equations 3-4 of 
\citealt*{beaetal2010}, and equation 1 of \citealt*{yamkim2014}) but
is applicable to a diverse range of systems, including for example the
the current Pluto-Charon system environment \citep[equation 3 of][]{porgru2014}
and the interstellar medium \citep[equation 3 of][]{howraf2014}.
However, the drag coefficient is, in general, not constant.  Equations 3.5-3.10 of
\cite{adaetal1976} show how the coefficient is actually a function of the Reynolds
number, the Mach number and the mean free path length of the gas molecules.
Recent applications which partly rely on results from physical experiments do not 
all use the same values for the different regimes 
(see equations 2-4 of \citealt*{garetal2004}, equations 15-20 of 
\citealt*{gibetal2012}, and equations 6-11 of \citealt*{lorbat2014}),
demonstrating slight disagreement.

We use the equations from \cite{garetal2004}, which can be expressed 
in our notation as

\begin{eqnarray}
\left(\frac{d\vec{v}}{dt}\right)_{\rm dr}&=\left\{ \begin{array}{ll} 
   \left(\frac{\rho_{\rm g}v_{\rm s}}{\rho R}\right)   \left( \vec{v}_{\rm g} - \vec{v} \right)
, & R\ll \zeta\\
   \left(\frac{\rho_{\rm g}B}{\rho R}\right)   \left( \vec{v}_{\rm g} - \vec{v} \right) 
                                              \left| \vec{v}_{\rm g} - \vec{v} \right|
, & R \gg \zeta
\end{array}\right .
\label{radvdrag}
\end{eqnarray}

\noindent{}where $\rho_{\rm g}$ is the density of the gas, $\zeta$ is the mean 
free path length of the gas, 
$\vec{v}_{\rm g} = (v_{\rm gx}, v_{\rm gy}, v_{\rm gz})$
is the velocity of the gas, $v_{\rm s}$ is the local sound speed, 
and

\begin{eqnarray}
B &=\left\{ \begin{array}{lll} 
   9\left[ \frac{6 R}{\zeta v_{\rm s}} \left| \vec{v}_{\rm g} - \vec{v} \right| \right]^{-1}
, & {\rm Re} \le 1\\
   9\left[ \frac{6 R}{\zeta v_{\rm s}} \left| \vec{v}_{\rm g} - \vec{v} \right| \right]^{-0.6}
, & 1 \le {\rm Re} \le 800 \\
0.165
, & {\rm Re} \ge 800
\end{array}\right .
\label{Resplit}
\end{eqnarray}

\noindent{}such that the Reynolds Number Re can be expressed as

\begin{equation}
{\rm Re} = \frac{6R}{\zeta v_{\rm s}} \left| \vec{v}_{\rm g} - \vec{v} \right|
.
\label{ReNum}
\end{equation}

The two pieces of equation (\ref{radvdrag}) refer to the {\it Epstein regime} (upper relation) and 
the {\it Stokes regime} (lower relation).  

Equations (\ref{radvdrag})-(\ref{ReNum}) should be applicable to GB star winds, just as they
would be for protoplanetary discs.  In their study of drag in GB systems, 
\cite{donetal2010} utilize one form of the drag equations in the Stokes regime, and for their
model suggest that the drag contribution is much smaller than gravity in the radial, but
not necessarily tangential, direction.  \cite{bonwya2010} instead compute a spiral-in timescale
due to GB stellar wind drag.

\subsubsection{The velocity lag}

A key quantity in equations (\ref{radvdrag})-(\ref{ReNum}) is the difference in velocity
between the gas and target.  This difference is also proportional to the Reynolds number
(equation \ref{ReNum}).  Therefore, let us explore this gradient in more detail.
Each component of both $\vec{v}$ and $\vec{v}_{\rm g}$ can be expressed in terms of the target's
orbital elements.  For $\vec{v}$, we use equations 5-6 of \cite{vereva2013a} and define
a rotation matrix $\mathbb{R}$ with components

\begin{eqnarray}
\mathbb{R}_{11} &=& \cos{\Omega} \cos{\omega} - \sin{\Omega} \sin{\omega} \cos{i}
,
\label{rotgen1}
\\
\mathbb{R}_{21} &=& \sin{\Omega} \cos{\omega} + \cos{\Omega} \sin{\omega} \cos{i}
,
\\
\mathbb{R}_{31} &=& \sin{\omega} \sin{i}
,
\\
\mathbb{R}_{12} &=& -\cos{\Omega} \sin{\omega} - \sin{\Omega} \cos{\omega} \cos{i} 
,
\\
\mathbb{R}_{22} &=& -\sin{\Omega} \sin{\omega} + \cos{\Omega} \cos{\omega} \cos{i} 
,
\\
\mathbb{R}_{32} &=& \cos{\omega} \sin{i}
,
\\
\mathbb{R}_{13} &=& \sin{\Omega} \sin{i}
,
\\
\mathbb{R}_{23} &=& -\cos{\Omega} \sin{i}
,
\\
\mathbb{R}_{33} &=& \cos{i}
,
\label{rotgen9} 
\end{eqnarray}

\noindent{}where $i$ is the inclination of target's orbit,
and $\Omega$ and $\omega$ are its longitude of ascending node
and argument of pericentre.  The target's velocity components are then

\begin{equation}
\vec{v} \equiv 
\mathbb{R}
\left( \begin{array}{c} 
            \frac{-na\sin{f}}{\sqrt{1-e^2}}  \\
            \frac{na\left(e+\cos{f}\right)}{\sqrt{1-e^2}}  \\
            0  \\
            \end{array}
            \right).
\label{velnor}
\end{equation}

The velocity of the gas (or wind) is assumed to be constant
and in the radial direction.  This velocity is dictated entirely by the star,
but shares the same space, and geometry, of the target. 
Consequently, we may express the components of $\vec{v}_{\rm g}$ at the target's location
in terms of the orbital parameters of the target as
\citep[equation 2.122 of][]{murder1999}

\begin{eqnarray}
v_{\rm gx} &=& \left| \vec{v}_{\rm g} \right|
\left[  
\cos{\Omega}\cos{\left(\omega + f\right)}
-
\sin{\Omega}\sin{\left(\omega + f\right)}\cos{i}
\right]
,
\label{gasgx}
\nonumber
\\
&&
\\
v_{\rm gy} &=& \left| \vec{v}_{\rm g} \right|
\left[  
\sin{\Omega}\cos{\left(\omega + f\right)}
+
\cos{\Omega}\sin{\left(\omega + f\right)}\cos{i}
\right]
,
\label{gasgy}
\nonumber
\\
&&
\\
v_{\rm gz} &=& \left| \vec{v}_{\rm g} \right|
\left[  
\sin{\left(\omega + f\right)}\sin{i}
\right]
.
\label{gasgz}
\end{eqnarray}

\noindent{}Equations (\ref{velnor})-(\ref{gasgy})
state that the gas velocity is in the radial direction.
However, we must carefully consider the
context when applying equations (\ref{gasgx})-(\ref{gasgz}).
For example, they would be incongruous if inserted
into averaged equations of motion.

The magnitude of $\vec{v}_{\rm g}$ is dictated
by the physical properties of the star. Because the 
launching mechanism of the wind is unknown \citep{blahof2012} and may vary as
a function of stellar evolution (see \citealt*{owocki2004} for a review),
we can make only an approximation.
We approximate the ejecta speed
as a function of the star's latitude 
as \citep{owocki2013} 

\begin{equation}
\left|\vec{v}_{\rm g}\right| = \sqrt{ \left( \frac{2GM}{R_{\star}(t)} \right) 
\left(1 - \frac{R_{\star}(t) V_{\rm rot}(t)^2}{G M} \cos^2{Z}  \right)
  }
,
\label{gasspeed}
\end{equation}

\noindent{}where $R_{\star}$ is the radius of the star, $Z$ is latitude, and $V_{\rm rot}$
is the rotational velocity of the star at the equator.  Time dependencies
are included to emphasize the potentially significant variance of these
quantities along the GB phases.  MS stars do also have winds, but for
fixed values of $R_{\star}$ and $V_{\rm rot}$.  WD stars have not yet been
observed to emit winds, and therefore in this context the concept of gas
drag through stellar winds does not exist and equation (\ref{gasspeed}) is not
applicable.






\subsubsection{The gas density, mean free path length, and  sound speed}

Before proceeding, we must make some approximations
about the gas density $\rho_{\rm g}$, the mean free path length 
(which we denote as $\zeta$), and the sound speed $v_{\rm s}$.  If the wind
is spherically symmetric so that $\rho_{\rm g}$ can be determined
from the mass loss rate, then we can write (equation 5 of \citealt*{donetal2010})

\begin{equation}
\frac{dM(t)}{dt} = 
4 \pi r^2 \rho_{\rm g} \left| \vec{v}_{\rm g} \right| 
.
\label{spsy}
\end{equation}

\noindent{}With a given mass loss rate, and a gas velocity given by
equation (\ref{gasspeed}), one can determine $\rho_{\rm g}$.

Given an expression for $\rho_{\rm g}$, how can we determine $\zeta$?
These quantities are related through

\begin{equation}
\zeta = \frac{\mu m_{\rm H}}{\theta \rho_{\rm g}}
,
\label{meanfree}
\end{equation}
\noindent{}where $\mu$ is the mean molecular mass
of the gas (or stellar wind),
$m_{\rm H}$ is the mass of a Hydrogen atom, and
$\theta$ is the collision cross-section of the gas
molecules. In equation (\ref{meanfree}), $m_{\rm H}$ is fixed and $\mu$ is on the
order of unity, so we need to bound the collisional cross-section $\theta$.
Based on the known radii of chemical elements, 
$10^{-20} {\rm m}^2 \lesssim \theta \lesssim 10^{-18} {\rm m}^2$.
Consequently, we make the approximate relation

\begin{equation}
\rho_{\rm g} \zeta \sim 10^{-8} {\rm kg}/{\rm m}^2
.
\end{equation}

\begin{figure}
\centerline{
\psfig{figure=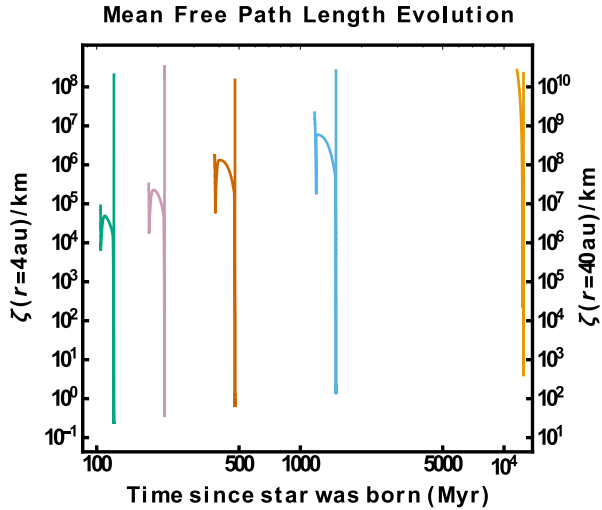,height=6.9cm} 
}
\caption{Evolution of the mean free path length of the stellar
wind for a fixed-in-space target across the giant branch phases 
of stellar evolution.
The target's physical and orbital parameters are $R = 1$ cm, $a = 5$ au 
(left $y$-axis), $a = 50$ au (right $y$-axis),
$e = 0.2$, $i = 0^{\circ}$, $\Omega = \omega = f = 0^{\circ}$.
The plot demonstrates how $\zeta$
undergoes changes of over 5 orders of magnitude, and achieves
a minimum at the tip of the AGB.
}
\label{zetaevo}
\end{figure}

\noindent{}By using this relation, we plot $\zeta$ in 
Figure \ref{zetaevo}.  The plot illustrates how at a fixed
point in space, $\zeta$ will vary by many orders of magnitude
during the GB phase. $\zeta$ is minimized at the tip of the AGB,
and scales with the square of the distance to the star.

The sound speed $v_{\rm s}$ is also unknown, and is a function
of the properties of the changing wind. \cite{owocki2014} suggests that
 $v_{\rm s}$ is a small fraction ($\sim 0.001$) of the escape velocity of the wind from
the star.  For simplicity and for lack of better constraints, we will assume that 
$v_{\rm s}$ is constant and given by this ratio.

\subsubsection{The Reynolds number}

\begin{figure}
\centerline{
\psfig{figure=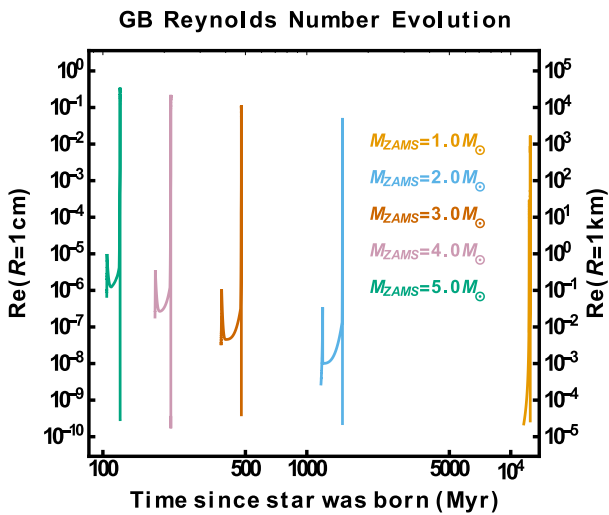,height=6.9cm} 
}
\caption{Evolution of the Reynolds number of the stellar
wind for a fixed-in-space target across the giant branch phases 
of stellar evolution.
The target's orbital parameters are $a = 5$ au, 
$e = 0.2$, $i = 0^{\circ}$, $\Omega = \omega = f = 0^{\circ}$.
The left and right $y$-axes respectively illustrate the
evolution for a $R = $ 1 cm pebble-like target and 
a $R = $ 1 km asteroid-like target.
The asteroid traverses all three of the Stokes flow
regimes (equation \ref{Resplit}).
}
\label{RayNu}
\end{figure}

Having now established methods or values for obtaining $\zeta$
and $v_{\rm s}$, we can attempt to estimate the Reynolds number
(equation \ref{ReNum}).  The Reynolds number is particularly important
because it defines the flow regime in which the target resides.
Figure \ref{RayNu} demonstrates how the variation in Re can be great enough
for a fixed point in space to be subject to several different flow regimes
throughout GB evolution.

\subsubsection{Comparing wind drag with radiation} \label{wdradcomp}

Having obtained some physical intuition for the variation of the Reynolds
number, we can now finally consider the acceleration due to wind drag.
In the same vein as we have defined $\beta$ and $\gamma$, here let us
define $\delta$ as the ratio of the perturbative acceleration due to
radiation to the perturbative acceleration due to wind drag.
In the Epstein regime,

\begin{eqnarray}
\delta &\lessapprox& 
\frac{3L\left(1 + e \cos{f} \right)^2}
{8\pi c a^2 \left(1 - e^2\right)^2 \rho_{\rm g} \left|v_{\rm g}\right|}
\left| \vec{v}_{\rm g} - \vec{v} \right|^{-1}
\nonumber
\\
&=&
\frac{3L}
{2 c \left| \vec{v}_{\rm g} - \vec{v} \right| }
\left(\frac{dM}{dt}\right)^{-1}
\label{deltaeqe}
\end{eqnarray}

\noindent{}whereas in the Stokes regime

\begin{eqnarray}
\delta &\lessapprox& 
\frac{3L\left(1 + e \cos{f} \right)^2}
{8\pi c a^2 \left(1 - e^2\right)^2 \rho_{\rm g} B}
\left| \vec{v}_{\rm g} - \vec{v} \right|^{-2}
\nonumber
\\
&=&
\frac{3L\left| \vec{v}_{\rm g} \right|}
{2 c B \left| \vec{v}_{\rm g} - \vec{v} \right|^2 }
\left(\frac{dM}{dt}\right)^{-1}
.
\label{deltaeqs}
\end{eqnarray}

\begin{figure}
\centerline{
\psfig{figure=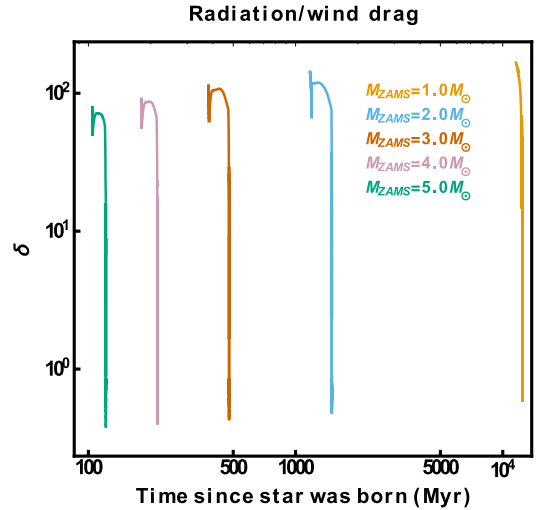,height=6.9cm} 
}
\caption{Relative importance of radiation versus wind
drag for a fixed-in-space target across the giant 
branch phases of stellar evolution.
The $R = 1$ cm target has orbital parameters of $a = 5$ au, 
$e = 0.2$, $i = 0^{\circ}$, and $\Omega = \omega = f = 0^{\circ}$.
However, the evolution of $\delta$ is largely insensitive
to any of these parameters and is instead largely determined
by the competition between the mass and luminosity evolution
of the star (equations \ref{deltaeqe} and \ref{deltaeqs}).
}
\label{deltaplot}
\end{figure}

The simplification we performed in equations (\ref{deltaeqe})-(\ref{deltaeqs})
holds only when one assumes the spherically symmetric wind of equation 
(\ref{spsy}).  In this approximation, $\delta$ is largely independent 
of the target's location or size, and instead is dependent almost entirely
on the properties of the star.  Figure \ref{deltaplot} illustrates
the evolution of $\delta$ for the stellar models that we adopted.
The curves suggest that the competition between radiation and wind
drag is close, and both should be considered together
in GB planetary studies.

\subsection{Comparison of all forces together}

\begin{figure*}
\centerline{\Large \bf Comparison of all effects}
\centerline{\Large \bf (dot-dashed $\equiv 1/\beta$, 
dashed $\equiv 1/\gamma$, dotted $\equiv 1/\delta$)}
\centerline{
\psfig{figure=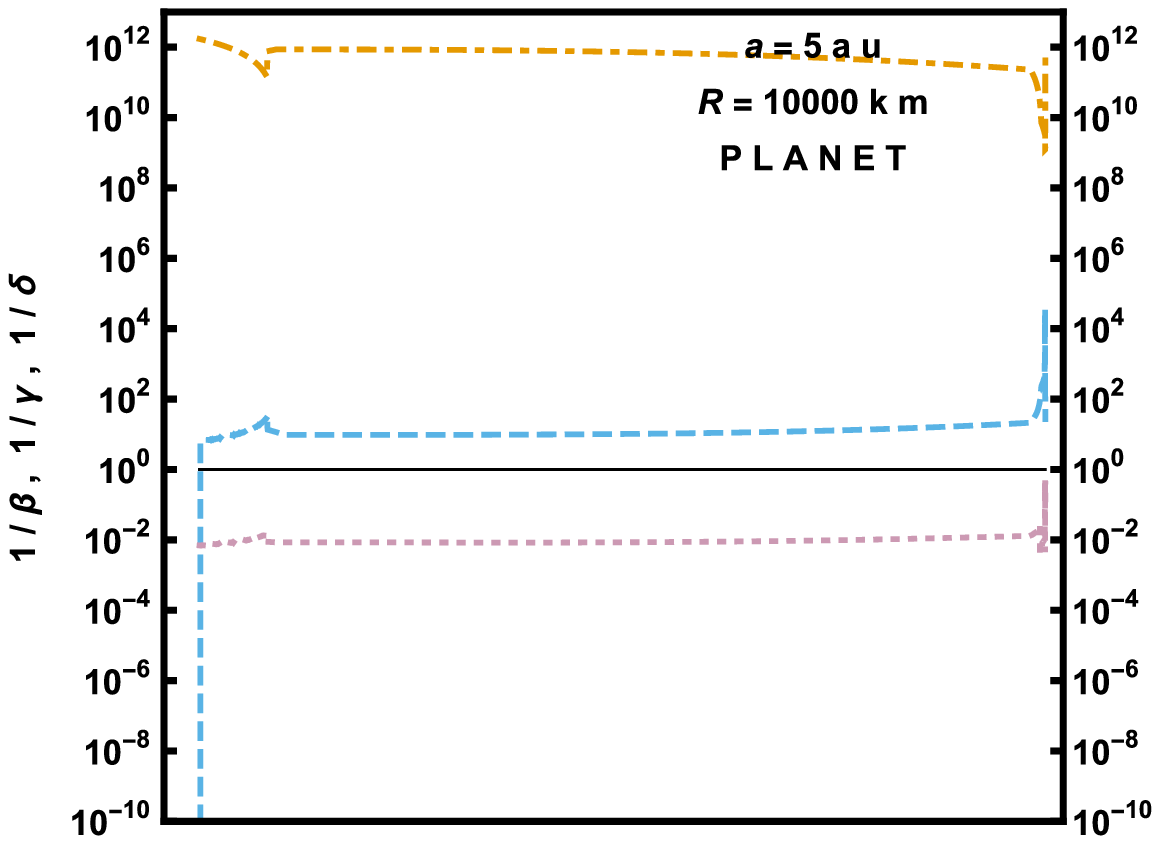,height=6.9cm} 
\psfig{figure=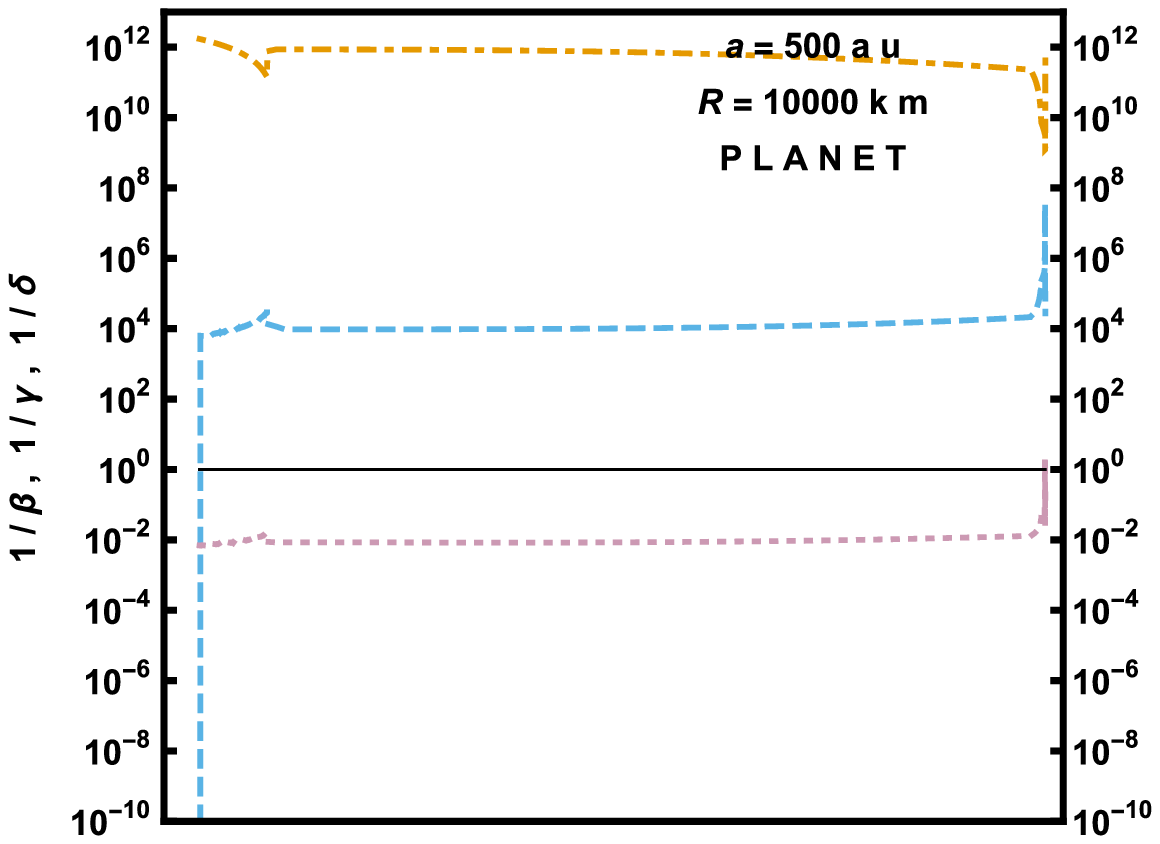,height=6.9cm}
}
\vspace{-1cm}
\centerline{
\psfig{figure=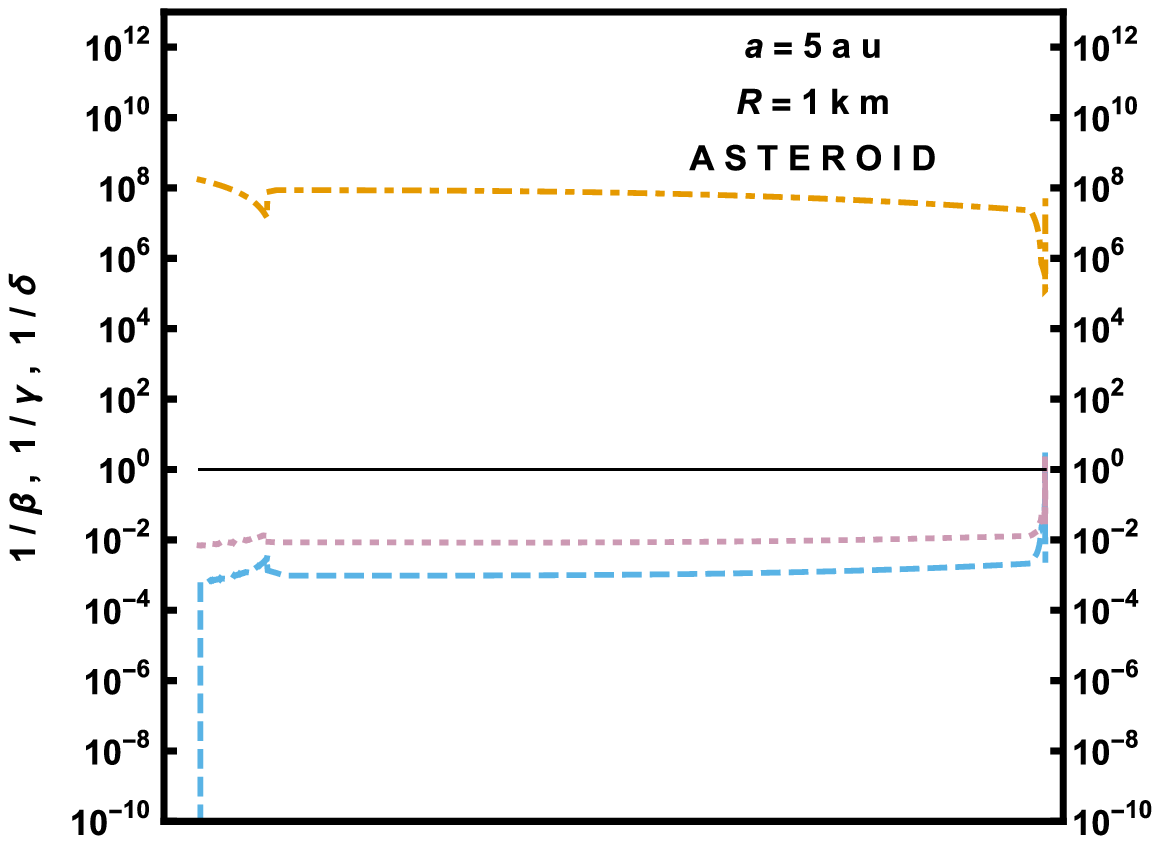,height=6.9cm} 
\psfig{figure=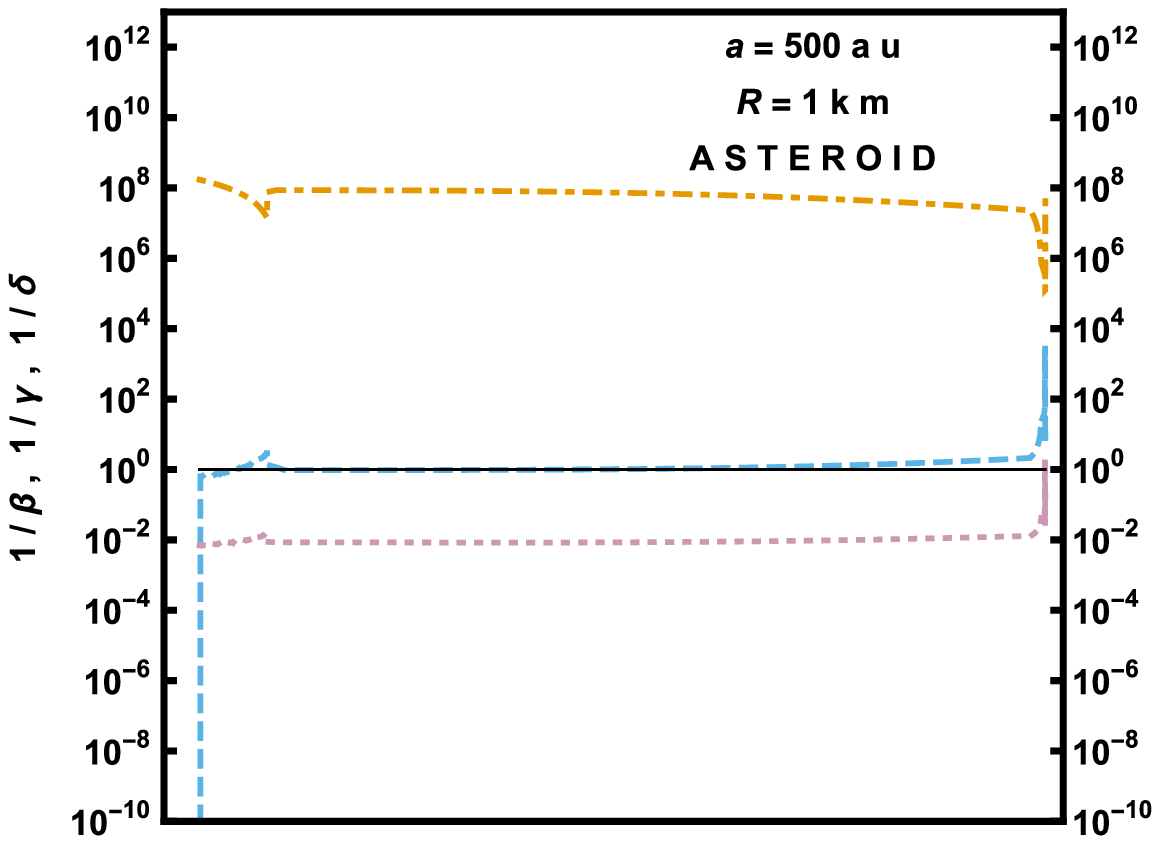,height=6.9cm}
}
\vspace{-1cm}
\centerline{
\psfig{figure=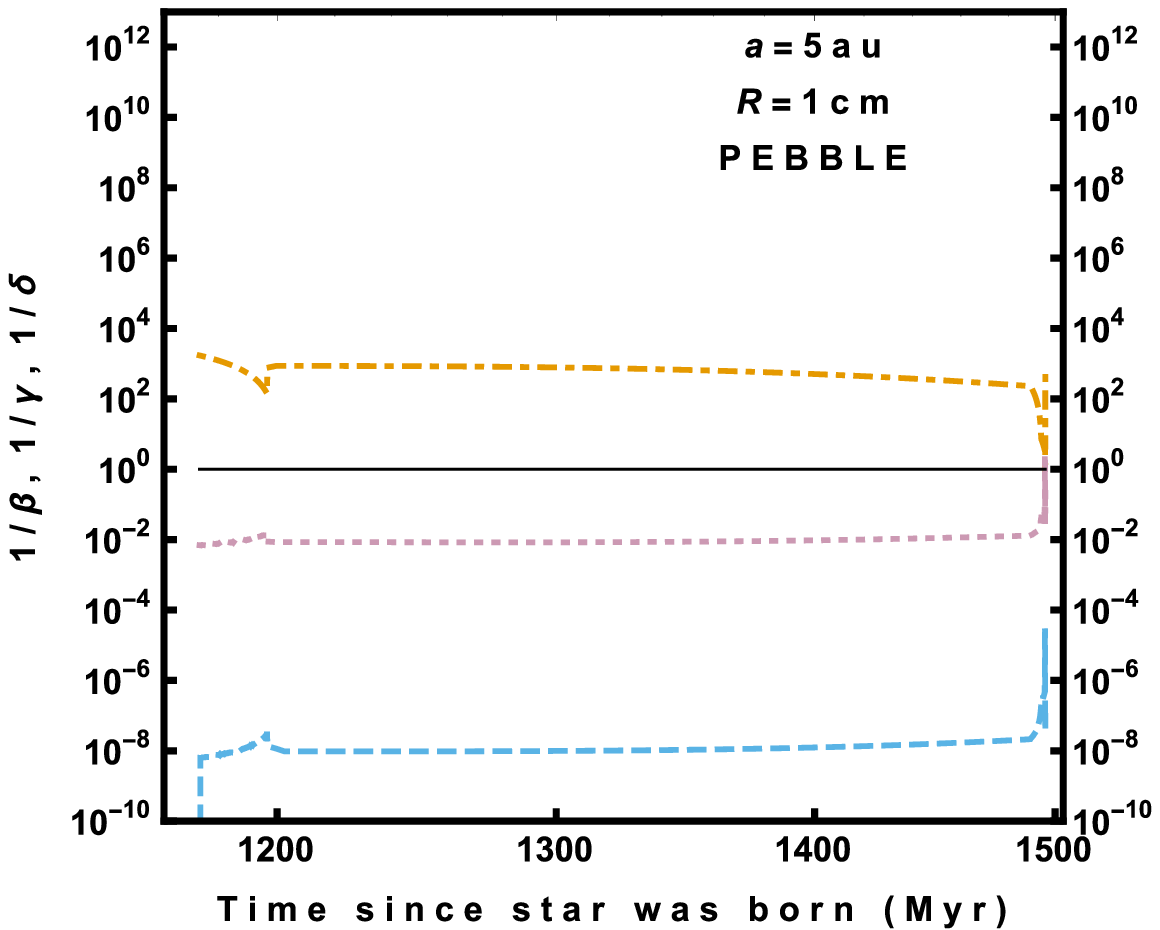,height=7.4cm} 
\psfig{figure=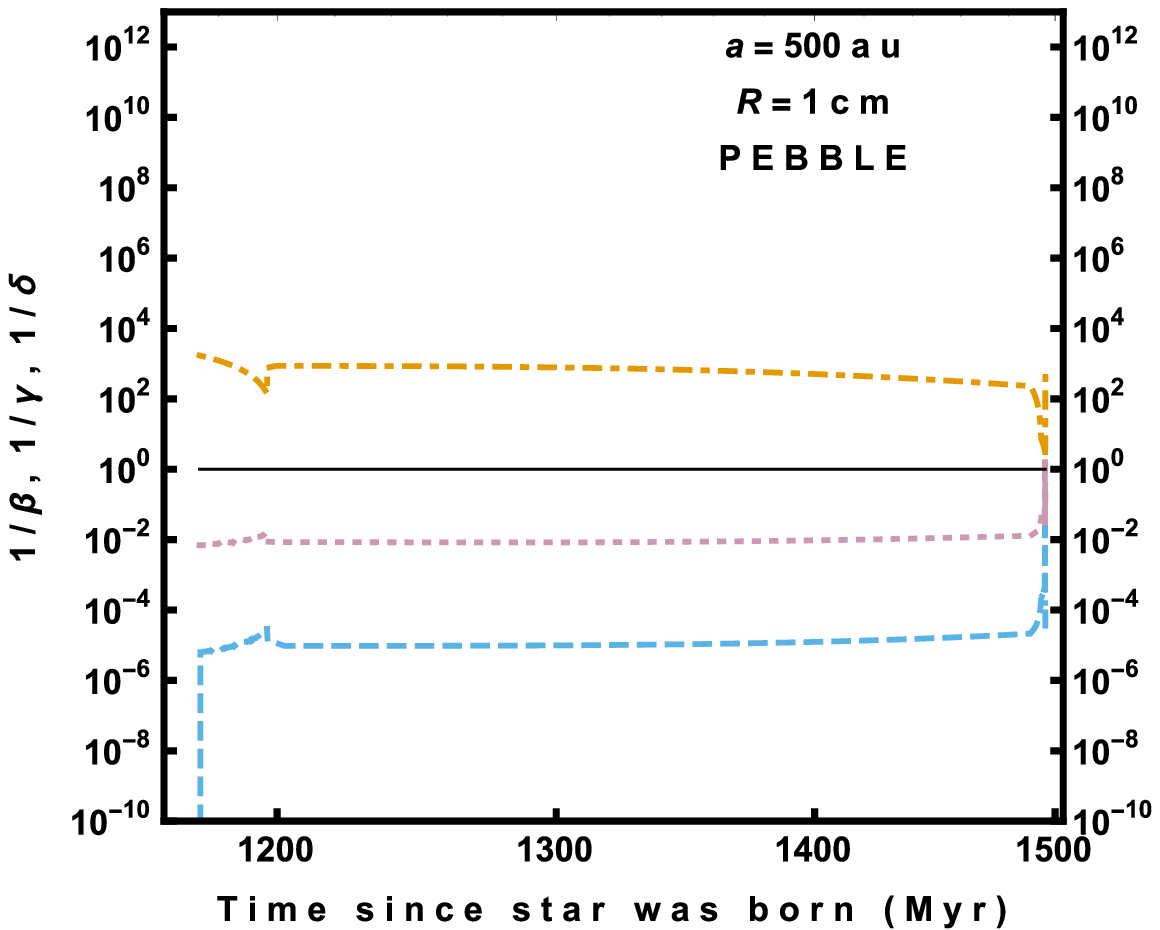,height=7.4cm}
}
\caption{Relative instantaneous acceleration strengths 
of gravity ($1/\beta$; orange dot-dashed lines),
mass loss ($1/\gamma$; blue dashed lines), and wind drag 
($1/\delta$; purple dotted lines), all compared to radiation
for $M_{\rm ZAMS} = 2M_{\odot}$ stars along the GB phases.  The relative
strengths and profile features for other progenitor stellar masses are similar.
The top, middle and bottom panels illustrate targets which are
a planet, asteroid and pebble, and the left and right panels
represent locations given by $a = 5$ au and $a = 500$ au assuming
the target is on a circular coplanar orbit.  This plot represents
just a snapshot in time and does not take into account 
potentially large long-term accumulations; detailed models should
instead consider all checked forces from Table \ref{Sumchart}.
}
\label{bigcompare}
\end{figure*}

Having defined the dimensionless ratios $\beta$, $\gamma$ and $\delta$,
we can now compare the strength of mass loss, gravity, wind drag
and radiation at given snapshots of time.  See Figure \ref{bigcompare}.
In all cases, radiation and wind drag are comparable in strength, and
comparable at all target radii sampled.  Recall that $\delta$ is
independent of $R$ everywhere except when the target is in the Stokes
regime and Re $\le 800$.  However, the pebble (with Re $\sim 10^{-9} - 10^{-2}$) 
and asteroid (with Re $\sim 10^{-4} - 10^{3}$) are both always in the Epstein 
regime here, and when the planet (with 
Re $\sim 10^{0} - 10^{7}$) is in the Stokes regime, usually Re $> 800$.
Consequently, changes in the purple dotted lines are almost indiscernible.
Expectedly, gravity dominates the evolution of planet-sized targets,
and mass loss becomes relatively more important the further one
moves away from the star.  These trends vary little for different
progenitor stellar masses.

Although broadly useful, this type of comparison does not
(i) characterize the evolution of individual orbital elements
(such as eccentricity), and (ii) fails to account for cumulative
effects.  Consequently, the middle panels suggest that gravity will
dominate the evolution of asteroids.  These plots do
not reveal that the long-term effect of the (radiative) Yarkovsky 
force can excite an asteroid's orbit to instability, as shown in
Section 4.3.4.  These plots are primarily useful for order of magnitude
estimates and short-term evolution; detailed evolutionary models 
should instead rely on Figure \ref{Sumchart}.

\section{Equations of motion in orbital elements}

The physical intuition gained in the last section from the comparison
of forces provides a foundation for the more precise treatment that
we supply here.  We seek to express the equations of motions due to 
wind drag and radiation entirely 
in terms of orbital elements through the formalism first mentioned in \cite{veretal2011} 
and developed in \cite{vereva2013a}.  Based on work
by \cite{efroimsky2005} and \cite{gurfil2007}, \cite{vereva2013a} delineated
how one can derive orbital-element-only equations from a two-body
problem with a perturbation expressed entirely in terms of Cartesian elements,
without any averaging.  This procedure assumes only boundedness of the orbits, and 
otherwise does not make any assumptions about the magnitude of the perturbation or 
the resulting orbital elements.

Among the benefits of this task is the
ability to then average over the result to create a new set of equations 
which can be integrated much more quickly and provide perspective on the target's
secular evolution.  Further, for the particular
case of perturbations due to radiation, we will demonstrate that in some
cases averaged leading-order terms vanish, helping to provide a better
assessment of the orbital changes than those from Figs. \ref{betaplot} and 
\ref{gammaplot}.  Henceforth, for ease of notation, we drop the explicit 
dependence of time on $M$ and $L$.

Unlike the other variables, the time evolution of the true anomaly need not be 
split up by type of force.  
The following relation holds true individually for mass loss, wind drag
and radiative forces, and together for the overall evolution

\begin{equation}
\frac{df}{dt} = \frac{n \left(1 + e \cos{f}\right)^2}{\left(1 - e^2\right)^{3/2}}
-\frac{d\omega}{dt}
-\cos{i} \frac{d\Omega}{dt}
.
\end{equation}

\noindent{}The true anomaly also represents the variable over which we average 
in order to obtain the secular equations.

\subsection{Mass loss equations}

First, for completeness and comparison, we repeat the known isotropic mass loss
equations.  Mass loss significantly changes the orbits of targets of
all sizes.

\subsubsection{Unaveraged equations}

The unaveraged equations of motion in orbital elements for isotropic mass loss
are \citep{omarov1962,hadjidemetriou1963,veretal2011}

\begin{eqnarray}
\left(\frac{da}{dt}\right)_{\rm ml} &=& -\frac{a \left(1 + e^2 + 2 e \cos{f}\right)}{1 - e^2}
\frac{1}{M+m} \frac{dM}{dt}
,
\label{mla}
\\
\left(\frac{de}{dt}\right)_{\rm ml} &=& -\left(e + \cos{f} \right)
\frac{1}{M+m} \frac{dM}{dt}
,
\\
\left(\frac{di}{dt}\right)_{\rm ml} &=& 0
,
\\
\left(\frac{d\Omega}{dt}\right)_{\rm ml} &=& 0
,
\\
\left(\frac{d\omega}{dt}\right)_{\rm ml} &=& -\frac{\sin{f}}{e} \frac{1}{M+m} \frac{dM}{dt}
.
\label{mlf}
\end{eqnarray}

\noindent{}The more complex anisotropic equations of motion \citep{veretal2013b} in fact more realistically
represent the motion, but, as previously suggested, they provide a negligible improvement on 
equations (\ref{mla}-\ref{mlf}) except in extreme cases.

\subsubsection{Averaged equations}

We obtain averages by performing the following integral for each variable
(here for averaged semimajor axis change due to mass loss)

\begin{equation}
\left\langle \left(\frac{da}{dt}\right)_{\rm ml} \right\rangle 
=
\frac{1}{2\pi}
\int_{0}^{2\pi} \left(\frac{da}{dt}\right)_{\rm ml}
\frac{\left(1-e^2\right)^{3/2}}{\left(1 + e \cos{f} \right)^2}
df.
\end{equation}

\noindent{}We find that all secular motions vanish except for the semimajor axis'

\begin{equation}
\left\langle \left(\frac{da}{dt}\right)_{\rm ml} \right\rangle = -\frac{a}{M+m}\frac{dM}{dt}
,
\label{MLavg}
\end{equation}

\begin{equation}
\left\langle \left(\frac{de}{dt}\right)_{\rm ml} \right\rangle 
= 
\left\langle \left(\frac{d\omega}{dt}\right)_{\rm ml} \right\rangle 
=
0
.
\label{MLavg2}
\end{equation}

\noindent{}The averaging assumes that $dM/dt$ is a constant and is much less than 
$df/dt$, such that $M(t)$ is also approximately constant on orbital
timescales.  Equations (\ref{MLavg})-(\ref{MLavg2}) represent an excellent approximation
for targets within hundreds of au \citep{veretal2011}.

\subsection{Wind drag equations}

Our derivation of the wind drag equations of motion in orbital
elements make no assumption about 
$\vec{v}_{\rm g} = (v_{\rm gx},v_{\rm gy},v_{\rm gz})$.  The gas
velocity here does not have to adhere to the prescriptions in 
equations (\ref{gasgx})-(\ref{gasspeed}) nor equation (\ref{spsy}).

\subsubsection{Auxiliary expressions}

In order to express our equations in a compact manner, we define
a standard set of auxiliary variables that appear regularly in studies of the perturbed
two-body problem
\citep{veretal2013b,vereva2013a,vereva2013b,veras2014a}.

\begin{eqnarray}
C_1 &\equiv& e \cos{\omega} + \cos{\left(f+\omega\right)}
\nonumber
\\
C_2 &\equiv& e \sin{\omega} + \sin{\left(f+\omega\right)}
\nonumber
\\
C_3 &\equiv& \cos{i} \sin{\Omega} \sin{\left(f+\omega\right)}
                   - \cos{\Omega} \cos{\left(f+\omega\right)}
\nonumber
\\ 
C_4 &\equiv& \cos{i} \cos{\Omega} \sin{\left(f+\omega\right)}
                   + \sin{\Omega} \cos{\left(f+\omega\right)}
\nonumber
\\
C_5 &\equiv& \left(3 + 4e \cos{f} + \cos{2f} \right) \sin{\omega}
\nonumber        
\\
\ \ \ \ &+&  
2 \left( e + \cos{f} \right) \cos{\omega} \sin{f}
\nonumber
\\
C_6 &\equiv& \left(3 + 4e \cos{f} + \cos{2f} \right) \cos{\omega}
\nonumber
\\        
\ \ \ \ &-& 2 \left( e + \cos{f} \right) \sin{\omega} \sin{f}
\nonumber
\\
C_7 &\equiv& \left(3 + 2e \cos{f} - \cos{2f} \right) \cos{\omega}
        + \sin{\omega} \sin{2f}
\nonumber
\\
C_8 &\equiv& \left(3  - \cos{2f} \right) \sin{\omega}
        - 2 \left(e + \cos{f} \right) \cos{\omega} \sin{f} 
\nonumber
\\
C_9 &\equiv& \left(3 + 2e \cos{f} - \cos{2f} \right) \sin{\omega}
        - \cos{\omega} \sin{2f}
.
\nonumber
\\
&&
\label{auxCs}
\end{eqnarray}

\noindent{}Additionally, in the following derivations we found that a commonly-occurring quantity is

\[
\mathcal{R} \equiv 
\bigg\lbrace
\left(v_{\rm gx} + \frac{an \left(C_2 \cos{\Omega} + C_1 \sin{\Omega} \cos{i}\right)}{\sqrt{1-e^2}}\right)^2
\]

\[
+
\left(v_{\rm gy} + \frac{an \left(-C_1 \cos{i}\cos{\Omega} + C_2 \sin{\Omega} \right)}{\sqrt{1-e^2}}\right)^2
\]

\begin{equation}
+
\left(v_{\rm gz} - \frac{an C_1 \sin{i}}{\sqrt{1-e^2}}\right)^2
\bigg\rbrace^{1/5}.
\label{mathcalReq}
\end{equation}

\subsubsection{Unaveraged equations for Epstein regime}

Recall that in the following derivations, we make no assumptions about $\vec{v}_{\rm g}$.
Consequently, from equation (\ref{radvdrag}) we obtain

\[
\left(\frac{da}{dt}\right)_{\rm dr} 
= 
\frac{2 v_{\rm s} \rho_{\rm g}}
{nR\rho\left(1-e^2\right) }
\bigg\lbrace
-an\left(1 + e^2 + 2 e \cos{f} \right)
\]

\[
\ \ \ \ + \sqrt{1-e^2}
\bigg[
v_{\rm gz} C_1 \sin{i} 
+ 
v_{\rm gy} \left(C_1 \cos{i} \cos{\Omega} - C_2 \sin{\Omega} \right)  
\]

\begin{equation}
\ \ \ \ -
v_{\rm gx} \left(C_1 \cos{i} \sin{\Omega} + C_2 \cos{\Omega} \right)  
\bigg]
\bigg\rbrace
,
\label{unavgepsi}
\end{equation}

\

\[
\left(\frac{de}{dt}\right)_{\rm dr} 
= 
-
\frac{v_{\rm s} \rho_{\rm g}}
{2anR\rho\left(1+e\cos{f}\right) }
\bigg\lbrace
4aen
\]

\[
\ \ \ \ +4an\cos{f}\left(1 + e^2 + e \cos{f} \right)
+ \sqrt{1-e^2}
\bigg[
-v_{\rm gz} C_6 \sin{i} 
\]

\[
\ \ \ \ + 
v_{\rm gy} \left(-C_6 \cos{i} \cos{\Omega} + C_5 \sin{\Omega} \right)  
\]

\begin{equation}
\ \ \ \ +
v_{\rm gx} \left(C_6 \cos{i} \sin{\Omega} + C_5 \cos{\Omega} \right)  
\bigg]
\bigg\rbrace
,
\end{equation}

\

\[
\left(\frac{di}{dt}\right)_{\rm dr} 
= 
\frac{\sqrt{1-e^2} v_{\rm s} \rho_{\rm g} \cos{\left(f+\omega\right)}}
{anR\rho\left(1+e\cos{f}\right) }
\]

\begin{equation}
\ \ \ \ \times \left[
v_{\rm gz} \cos{i} 
- 
v_{\rm gy} \sin{i} \cos{\Omega}  
+
v_{\rm gx} \sin{i} \sin{\Omega}  
\right]
,
\end{equation}

\

\[
\left(\frac{d\Omega}{dt}\right)_{\rm dr} 
= 
\frac{\sqrt{1-e^2} v_{\rm s} \rho_{\rm g} \sin{\left(f+\omega\right)}}
{anR\rho\left(1+e\cos{f}\right) }
\]

\begin{equation}
\ \ \ \ \times \left[
v_{\rm gz} \cot{i} 
- 
v_{\rm gy} \cos{\Omega}  
+
v_{\rm gx} \sin{\Omega}  
\right]
,
\end{equation}

\

\[
\left(\frac{d\omega}{dt}\right)_{\rm dr} 
= 
-
\frac{v_{\rm s} \rho_{\rm g}}
{4anR\rho e\sqrt{1-e^2}\left(1+e\cos{f}\right) }
\bigg\lbrace
\]

\[
\ \ \ \ 8an\sqrt{1-e^2}\left(1 + e \cos{f} \right) \sin{f} - 2\left(1 - e^2 \right)
\]

\[
\ \ \ \ \times \bigg[
-v_{\rm gz} \left[C_9 \sin{i} + 2 e \cos{i} \cot{i} \sin{\left(f + \omega\right)}   \right]
\]

\[
\ \ \ \ - 
v_{\rm gy} \left(C_8 \cos{i} \cos{\Omega} + C_7 \sin{\Omega} \right)  
\]

\begin{equation}
\ \ \ \ +
v_{\rm gx} \left(-C_7 \cos{\Omega} + C_8 \cos{i} \sin{\Omega} \right)  
\bigg]
\bigg\rbrace
.
\label{unavgepsf}
\end{equation}

Equations (\ref{unavgepsi})-(\ref{unavgepsf}) reveal that all the orbital elements vary
along a single orbit due to Epstein drag, even if the orbit is initially eccentric
or circular.  This fact remains true even if spherical symmetry is imposed.

\subsubsection{Unaveraged equations for Stokes regime, Re $\le 1$}

The orbital element evolution equations of motion in this regime
are equivalent to those in Section 4.2.2, but with right-hand-sides
multiplied by a factor
of $3\zeta/2R$.  Recall that targets will rarely find themselves
in this regime, and must be close to the star to do so.

\subsubsection{Unaveraged equations for Stokes regime, $1~\le~$~Re~$\le~800$}

The equations of motion in this subregime have a similar form to those
in Section 4.2.2, but are scaled differently.

\[
\left(\frac{da}{dt}\right)_{\rm dr} 
= 
-\frac{3 \cdot 6^{2/5} \zeta^{3/5}  v_{\rm s}^{3/5} \rho_{\rm g} \mathcal{R}}
{\sqrt{a} n \rho \left(1 - e^2 \right) R^{8/5}   }
\]

\[
\ \ \ \ \times\bigg\lbrace
n a^{3/2} \left(1 + e^2 + 2 e \cos{f} \right) 
+
a\sqrt{1-e^2}
\]

\[
\ \ \ \ \bigg[
-v_{\rm gz} C_1 \sin{i} 
+ 
v_{\rm gy} \left(-C_1 \cos{i} \cos{\Omega} + C_2 \sin{\Omega} \right)  
\]

\begin{equation}
\ \ \ \ +
v_{\rm gx} \left(C_1 \cos{i} \sin{\Omega} + C_2 \cos{\Omega} \right)  
\bigg]
\bigg\rbrace
,
\end{equation}

\

\[
\left(\frac{de}{dt}\right)_{\rm dr} 
= 
-\frac{3^{7/5} \zeta^{3/5}  v_{\rm s}^{3/5} \rho_{\rm g} \mathcal{R}}
{2^{8/5}a^{3/2} n \rho \left(1 + e \cos{f} \right) R^{8/5}   }
\]

\[
\ \ \ \ \times\bigg\lbrace
2 a^{3/2} n
\left[
2 \cos{f} \left(1 + e^2\right)
+
e \left(3 + \cos{2f}  \right)
\right]
+
a\sqrt{1-e^2}
\]

\[
\ \ \ \ \bigg[
-v_{\rm gz} C_6 \sin{i} 
+ 
v_{\rm gy} \left(-C_6 \cos{i} \cos{\Omega} + C_5 \sin{\Omega} \right)  
\]

\begin{equation}
\ \ \ \ +
v_{\rm gx} \left(C_6 \cos{i} \sin{\Omega} + C_5 \cos{\Omega} \right)  
\bigg]
\bigg\rbrace
,
\end{equation}

\

\[
\left(\frac{di}{dt}\right)_{\rm dr} 
= 
\frac{3^{7/5} \zeta^{3/5}  v_{\rm s}^{3/5} \rho_{\rm g} \mathcal{R} \sqrt{1-e^2}\cos{\left(f+\omega\right)}}
{2^{3/5}a n \rho \left(1 + e \cos{f} \right) R^{8/5}   }
\]

\begin{equation}
\ \ \ \ \times\bigg\lbrace
v_{\rm gz} \cos{i}
-
v_{\rm gy} \sin{i}\cos{\Omega}
+
v_{\rm gx} \sin{i}\sin{\Omega}
\bigg\rbrace,
\end{equation}

\

\[
\left(\frac{d\Omega}{dt}\right)_{\rm dr} 
= 
\frac{3^{7/5} \zeta^{3/5}  v_{\rm s}^{3/5} \rho_{\rm g} \mathcal{R} \sqrt{1-e^2}\sin{\left(f+\omega\right)}}
{2^{3/5}a n \rho \left(1 + e \cos{f} \right) R^{8/5}   }
\]

\begin{equation}
\ \ \ \ \times\bigg\lbrace
v_{\rm gz} \cot{i}
-
v_{\rm gy} \cos{\Omega}
+
v_{\rm gx} \sin{\Omega}
\bigg\rbrace,
\end{equation}

\

\[
\left(\frac{d\omega}{dt}\right)_{\rm dr} 
= 
-\frac{3^{7/5} \zeta^{3/5}  v_{\rm s}^{3/5} \rho_{\rm g} \mathcal{R}}
{2^{8/5}a n e \sqrt{1-e^2} \rho \left(1 + e \cos{f} \right) R^{8/5}   }
\]

\[
\ \ \ \ \times\bigg\lbrace
4 a n \sqrt{1-e^2} \sin{f} \left(1 + e \cos{f} \right)  
- 
\left(1 - e^2 \right)
\]

\[
\ \ \ \ \times\bigg[
-v_{\rm gz} \left( C_9 \sin{i} + 2 e \cos{i} \cot{i} \sin{\left(f + \omega \right)}   \right) 
\]

\[
\ \ \ \ - 
v_{\rm gy} \left(C_8 \cos{i} \cos{\Omega} + C_7 \sin{\Omega} \right)  
\]

\begin{equation}
\ \ \ \ +
v_{\rm gx} \left(C_8 \cos{i} \sin{\Omega} - C_7 \cos{\Omega} \right)  
\bigg]
\bigg\rbrace
.
\end{equation}

\subsubsection{Unaveraged equations for Stokes regime, $800~\le$~Re}

In this subregime, $B$ is a constant.  Let $B = 0.165 \equiv \mathcal{K}$.
Although the equations simplify and $(di/dt)_{\rm dr}$ 
and $(d\Omega/dt)_{\rm dr}$ vanish, the
remaining equations are still not generally solvable analytically.

The full equations are

\[
\left(\frac{da}{dt}\right)_{\rm dr} 
= 
-
\frac
{2 \mathcal{K} \rho_{\rm g} \mathcal{R}^{5/2}}
{n R \rho \left(1 - e^2 \right)}
\bigg\lbrace
an \left(1 + e^2 + 2 e \cos{f} \right)
\]

\[
\ \ \ \ +
\sqrt{1-e^2}
\bigg[
-
v_{\rm gz} C_1 \sin{i}
+
v_{\rm gy} \left(-C_1 \cos{\Omega} \cos{i} + C_2 \sin{\Omega} \right)
\]

\begin{equation}
\ \ \ \ +
v_{\rm gx} \left(C_1 \sin{\Omega} \cos{i} + C_2 \cos{\Omega} \right)
\bigg]
\bigg\rbrace
,
\end{equation}

\

\[
\left(\frac{de}{dt}\right)_{\rm dr} 
= 
-
\frac
{\mathcal{K} \rho_{\rm g} \mathcal{R}^{5/2}}
{2 n a R \rho \left(1 + e \cos{f} \right)}
\]

\[
\ \ \ \ \times\bigg\lbrace
2an \left[ 3e + 2\left(1 + e^2 \right) \cos{f} + e \cos{\left(2f\right)}  \right]
\]

\[
\ \ \ \ +
\sqrt{1-e^2}
\bigg[
-
v_{\rm gz} C_6 \sin{i}
\]

\[
\ \ \ \ +
v_{\rm gy} \left(-C_6 \cos{\Omega} \cos{i} + C_5 \sin{\Omega} \right)
\]

\begin{equation}
\ \ \ \ +
v_{\rm gx} \left(C_6 \sin{\Omega} \cos{i} + C_5 \cos{\Omega} \right)
\bigg]
\bigg\rbrace,
\end{equation}

\

\[
\left(\frac{di}{dt}\right)_{\rm dr} 
= 
\frac
{\mathcal{K} \rho_{\rm g} \mathcal{R}^{5/2} \sqrt{1-e^2} \cos{\left(f + \omega \right)}}
{a n R \rho \left(1 + e \cos{f} \right)}
\]

\begin{equation}
\ \ \ \ \times\bigg\lbrace
v_{\rm gz} \cos{i} - v_{\rm gy} \sin{i} \cos{\Omega} + v_{\rm gx} \sin{i} \sin{\Omega}
\bigg\rbrace,
\end{equation}

\

\[
\left(\frac{d\Omega}{dt}\right)_{\rm dr} 
= 
\frac
{\mathcal{K} \rho_{\rm g} \mathcal{R}^{5/2} \sqrt{1-e^2} \sin{\left(f + \omega \right)}}
{a n R \rho \left(1 + e \cos{f} \right)}
\]

\begin{equation}
\ \ \ \ \times\bigg\lbrace
v_{\rm gz} \cot{i} - v_{\rm gy} \cos{\Omega} + v_{\rm gx} \sin{\Omega}
\bigg\rbrace,
\end{equation}

\

\[
\left(\frac{d\omega}{dt}\right)_{\rm dr} 
= 
\frac
{\mathcal{K} \rho_{\rm g} \mathcal{R}^{5/2}}
{2 n a R \rho e \sqrt{1-e^2} \left(1 + e \cos{f} \right)}
\]

\[
\ \ \ \ \times\bigg\lbrace
-an \sqrt{1-e^2} \left( C_2 C_7 - C_1 C_9 \right)
-
\left(1-e^2\right)
\]

\[
\ \ \ \ \bigg[
v_{\rm gz} \left( C_9 \sin{i} +  2 e \cos{i} \cot{i} \sin{\left(f + \omega \right)}  \right)
\]

\[
\ \ \ \ -
v_{\rm gy} \left[-\cos{i} \cos{\Omega} \left(C_9 - 2 e \sin{\left( f+\omega \right) } \right) - C_7 \sin{\Omega} \right]
\]

\begin{equation}
\ \ \ \ +
v_{\rm gx} \left[-\cos{i} \sin{\Omega} \left(C_9 - 2 e \sin{\left( f+\omega \right) } \right) + C_7 \cos{\Omega} \right]
\bigg]
\bigg\rbrace
.
\label{finalsto}
\end{equation}

\subsubsection{Averaged equations}

The unaveraged equations demonstrated exactly how the osculating
orbital elements change at a given point in time, and most practically,
along a single orbit.  Averaging over these equations would
require additional assumptions, which we do not make here.  One can,
for example, assume a spherically symmetric wind (as in equation \ref{spsy})
and then some particular analytical prescription for stellar mass loss to treat
the variation of $\vec{v}_{\rm g}$.  However, even with those assumptions, 
performing averaging over variables such as $\mathcal{R}$ (applicable in the Stokes 
regime, with Re $> 1$) becomes computationally prohibative.

\subsection{Radiation equations} \label{radsec}

The equations of motion due to radiative effects 
(from equation \ref{finaldvdtp}) are 
too complex to be reasonably expressed
in osculating orbital elements
due to the Yarkovsky contribution.  Because the diurnal Yarkovsky acceleration contains an
explicit dependence on $s$, one must also treat the evolution of the target's
spin.  We can do so through equation (\ref{sdef}) only if we assume a slow
rate of change.
Further, because the seasonal Yarkovsky
contribution is an explicit function of the specific angular momentum through
equations (\ref{rothxi}), (\ref{r1hmat}), (\ref{r2hmat}) and (\ref{xiang}),
the final result is unmanageable even if the diurnal contribution is neglected.

Nevertheless, our perturbation method is sufficient for the purposes of this paper,
as we aim to place bounds on the motion and not model any particular system.
Consequently, when deriving the perturbation equations below, we assume that
the matrix elements of $\mathbb{Q}$ are independent of position and velocity.

\subsubsection{Auxiliary expressions}

Even with this assumption, the final resulting equations are long.
However, we have discovered that they can be expressed compactly 
by using a smart choice of auxiliary variables.  First we 
define two additional $C$ variables that represent commonly
occurring quantities in the equations of motion of our problem.

\begin{eqnarray}
C_{10} &\equiv& -3 e \sin{\omega} - 2 \sin{\left(f+\omega\right)} + e \sin{\left(2f+\omega\right)}
,
\nonumber
\\
C_{11} &\equiv& -3 e \cos{\omega} - 2 \cos{\left(f+\omega\right)} + e \cos{\left(2f+\omega\right)}
.
\nonumber
\\
&&
\label{C10C11}
\end{eqnarray}

Also, we found that one quantity which appears in all of the terms that are 
proportional to $1/c$ is

\[
\vec{D}_c \equiv 
-C_3
\left( \begin{array}{c} 
            \mathbb{Q}_{11}  \\
            \mathbb{Q}_{21}  \\
            \mathbb{Q}_{31}  \\
            \end{array}
            \right)
+C_4
\left( \begin{array}{c} 
            \mathbb{Q}_{12}  \\
            \mathbb{Q}_{22}  \\
            \mathbb{Q}_{32}  \\
            \end{array}
            \right) 
\]

\begin{equation}
\ \ \
+\sin{i}\sin{\left(f+\omega\right)}
\left( \begin{array}{c} 
            \mathbb{Q}_{13}  \\
            \mathbb{Q}_{23}  \\
            \mathbb{Q}_{33}  \\
            \end{array}
            \right)
.
\label{auxDc}
\end{equation}

\noindent{}A quantity appearing in all the terms that are proportional to $1/c^2$ is

\[
\vec{D}_{c^2} \equiv 
-\left(C_{10} \cos{\Omega} + C_{11} \cos{i} \sin{\Omega} \right)
\left( \begin{array}{c} 
            \mathbb{Q}_{11}  \\
            \mathbb{Q}_{21}  \\
            \mathbb{Q}_{31}  \\
            \end{array}
            \right)
\]

\[
\ \ \
-\left(C_{10} \sin{\Omega} - C_{11} \cos{i} \cos{\Omega} \right)
\left( \begin{array}{c} 
            \mathbb{Q}_{12}  \\
            \mathbb{Q}_{22}  \\
            \mathbb{Q}_{32}  \\
            \end{array}
            \right) 
\]

\begin{equation}
\ \ \ 
+C_{11} \sin{i}
\left( \begin{array}{c} 
            \mathbb{Q}_{13}  \\
            \mathbb{Q}_{23}  \\
            \mathbb{Q}_{33}  \\
            \end{array}
            \right) 
\end{equation}

\noindent{}Terms specific to the orbital element expressions are 

\begin{equation}
\vec{D}_{a}
\equiv
\left( \begin{array}{c} 
            -C_2 \cos{\Omega} - C_1 \sin{\Omega} \cos{i} \\
            -C_2 \sin{\Omega} + C_1 \cos{\Omega} \cos{i} \\
            C_1 \sin{i}  \\
            \end{array}
            \right)
,
\end{equation}

\begin{equation}
\vec{D}_{e}
\equiv
\left( \begin{array}{c} 
            -C_5 \cos{\Omega} - C_6 \sin{\Omega} \cos{i} \\
            -C_5 \sin{\Omega} + C_6 \cos{\Omega} \cos{i} \\
            C_6 \sin{i}  \\
            \end{array}
            \right)
,
\end{equation}

\begin{equation}
\vec{D}_{\Omega}
\equiv
\left( \begin{array}{c} 
            \sin{\Omega}  \\
           -\cos{\Omega}  \\
            \cot{i}  \\
            \end{array}
            \right)
,
\end{equation}

\begin{equation}
\vec{D}_{i} = \sin{i} \vec{D}_{\rm \Omega},
\end{equation}

\begin{equation}
\vec{D}_{\omega} 
\equiv
\left( \begin{array}{c} 
            -C_7 \cos{\Omega} + C_8 \sin{\Omega} \cos{i} \\
            -C_7 \sin{\Omega} - C_8 \cos{\Omega} \cos{i} \\
            -C_9 \csc{i} + C_8 \cos{i} \cot{i}  \\
            \end{array}
            \right)
.
\label{auxDomega}
\end{equation}

\subsubsection{Unaveraged equations}

Equipped with the above auxiliary variables, the
complete final equations of motion entirely in orbital elements are

\[
\left(\frac{da}{dt}\right)_{\rm ra} 
=
\left(\frac{1}{c}\right)
\frac{A L \left(1+e \cos{f} \right)^2}{2\pi m n a^2 \left(1 - e^2\right)^{5/2}}
\left( \vec{D}_c \cdot \vec{D}_a \right)
\]

\begin{equation}
\ \ \ \ \
+ 
\left(\frac{1}{c^2}\right)
\frac{A L \left(1+e \cos{f} \right)^2}{4\pi m a \left(1 - e^2\right)^{3}}
\left( \vec{D}_{c^2} \cdot \vec{D}_a \right)
,
\label{firstDeq}
\end{equation}

\

\

\[
\left(\frac{de}{dt}\right)_{\rm ra}
=
\left(\frac{1}{c}\right)
\frac{A L \left(1+e \cos{f} \right)}{8\pi m n a^3 \left(1 - e^2\right)^{3/2}}
\left( \vec{D}_c \cdot \vec{D}_e \right)
\]

\begin{equation}
\ \ \ \ \
+ 
\left(\frac{1}{c^2}\right)
\frac{A L \left(1+e \cos{f} \right)}{16\pi m a^2 \left(1 - e^2\right)^2}
\left( \vec{D}_{c^2} \cdot \vec{D}_e \right)
,
\end{equation}

\

\

\[
\left(\frac{di}{dt}\right)_{\rm ra} 
=
\left(\frac{1}{c}\right)
\frac{A L \left(1+e \cos{f} \right)\cos{\left(f+\omega\right)}}{4\pi m n a^3 \left(1 - e^2\right)^{3/2}}
\left( \vec{D}_c \cdot \vec{D}_i \right)
\]

\begin{equation}
\ \ \ \ \
+ 
\left(\frac{1}{c^2}\right)
\frac{A L \left(1+e \cos{f} \right)\cos{\left(f+\omega\right)}}{8\pi m a^2 \left(1 - e^2\right)^2}
\left( \vec{D}_{c^2} \cdot \vec{D}_i \right)
,
\end{equation}

\

\

\[
\left(\frac{d\Omega}{dt}\right)_{\rm ra}
=
\left(\frac{1}{c}\right)
\frac{A L \left(1+e \cos{f} \right)\sin{\left(f+\omega\right)}}{4\pi m n a^3 \left(1 - e^2\right)^{3/2}}
\left( \vec{D}_c \cdot \vec{D}_{\Omega} \right)
\]

\begin{equation}
\ \ \ \ \
+ 
\left(\frac{1}{c^2}\right)
\frac{A L \left(1+e \cos{f} \right)\sin{\left(f+\omega\right)}}{8\pi m a^2 \left(1 - e^2\right)^2}
\left( \vec{D}_{c^2} \cdot \vec{D}_{\Omega} \right)
,
\end{equation}

\

\

\[
\left(\frac{d\omega}{dt}\right)_{\rm ra}
=
\left(\frac{1}{c}\right)
\frac{A L \left(1+e \cos{f} \right)}{8\pi m n a^3 e \left(1 - e^2\right)^{3/2}}
\left( \vec{D}_c \cdot \vec{D}_{\omega} \right)
\]

\begin{equation}
\ \ \ \ \
+ 
\left(\frac{1}{c^2}\right)
\frac{A L \left(1+e \cos{f} \right)}{16\pi m a^2 e \left(1 - e^2\right)^2}
\left( \vec{D}_{c^2} \cdot \vec{D}_{\omega} \right)
.
\label{lastDeq}
\end{equation}

\noindent{}Equations (\ref{firstDeq})-(\ref{lastDeq}) are convenient 
because all of the $\vec{D}$ auxiliary variables, which are on the order 
of unity, are isolated.  They also show that the amplitudes of the 
orbital element rates of changes are independent of $i$, and 
$\Omega$.  Further, for many purposes, the $1/c^2$ term may be neglected.

\subsubsection{Unaveraged equations with no Yarkovsky}

When the Yarkovsky effect is ``off'', which typically occurs for
targets with diameters less than 1 cm-1 m,
a significant simplification occurs.  Now, $\mathbb{Q}$ is a diagonal
matrix, and we can convert that into a scalar.  Consequently, equation
(\ref{finaldvdtp}) becomes

\begin{equation}
\left(\frac{d\vec{v}}{dt}\right)_{\rm ra}\bigg|_{\mathbb{Y}=0}
=
\frac
{AL \left(Q_{\rm abs} + Q_{\rm ref}\right)}
{4\pi m c r^2}
\vec{\iota}
.
\label{finaldvdtp2}
\end{equation}

\noindent{}Subsequently, the equations of motion simplify 
to the following relations

\[
\left( \frac{da}{dt} \right)_{\rm ra}\bigg|_{\mathbb{Y}=0}
=
\frac{AL\left(Q_{\rm abs} + Q_{\rm ref}\right)
\left(1+e\cos{f}\right)^2}{2\pi m\left(1-e^2\right)^3}
\]

\[
\ \ \ \ \times
\bigg[
\frac
{e\sqrt{1-e^2} \sin{f} }
{n a^2}
\left(\frac{1}{c}\right)
\]

\begin{equation}
\ \ \ \
+
\frac
{-2 -3e^2 - 4e\cos{f} + e^2 \cos{\left(2f\right)} }
{2 a}
\left(\frac{1}{c^2}\right)
\bigg]
,
\end{equation}

\[
\left(
\frac{de}{dt}
\right)_{\rm ra}\bigg|_{\mathbb{Y}=0}
=
\frac{AL\left(Q_{\rm abs} + Q_{\rm ref}\right)
\left(1+e\cos{f}\right)^2}{4\pi m\left(1-e^2\right)^{3/2}}
\]

\begin{equation}
\ \ \ \
\times
\bigg[
\frac
{\sin{f} }
{n a^3}
\left(\frac{1}{c}\right)
+
\frac
{- 4\cos{f} + e \left(\cos{\left(2f\right)} - 5\right)}
{2 a^2 \sqrt{1-e^2}}
\left(\frac{1}{c^2}\right)
\bigg]
,
\end{equation}

\begin{equation}
\left(
\frac{di}{dt}
\right)_{\rm ra}\bigg|_{\mathbb{Y}=0}
=
\left(
\frac{d\Omega}{dt}
\right)_{\rm ra}\bigg|_{\mathbb{Y}=0}
=
0
,
\end{equation}

\[
\left( \frac{d\omega}{dt} \right)_{\rm ra}\bigg|_{\mathbb{Y}=0}
=
\left( \frac{d\varpi}{dt} \right)_{\rm ra}\bigg|_{\mathbb{Y}=0}
=
\]

\[
\ \ \ \
-
\frac{AL\left(Q_{\rm abs} + Q_{\rm ref}\right)
\left(1+e\cos{f}\right)^2}{4\pi m e\left(1-e^2\right)^{3/2}}
\]

\begin{equation}
\ \ \ \
\times
\bigg[
\frac
{\cos{f} }
{n a^3}
\left(\frac{1}{c}\right)
+
\frac
{\sin{f} \left(2 - e\cos{f}\right)}
{a^2 \sqrt{1-e^2}}
\left(\frac{1}{c^2}\right)
\bigg]
.
\label{domegadt}
\end{equation}

For pebbles which are subject to intense radiation, equations
(\ref{finaldvdtp2})-(\ref{domegadt}) describe their evolution
due to this perturbation exactly.  For asteroids, however, 
equations (\ref{firstDeq})-(\ref{lastDeq}) should be used
instead.  For particular models which require a detailed
analysis of the perturbation at the pericentre or the stationary
points of the motion along an orbit, equations (\ref{finaldvdtp2})-(\ref{domegadt})
are compact enough to allow for an analysis similar to the
one by \cite{veras2014b}.

\subsubsection{Averaged equations}

Now we return to the general equations in 
(\ref{firstDeq})-(\ref{lastDeq}) and average over them.
Integrating these equations yields long
expressions.  We write out these expressions in the appendix
(with the $1/c$ term only) because they can be cast in a revealing form.

Just by inspection of equations (\ref{dadtavgYark})-(\ref{domegadtavgYark}), one
can see that when the Yarkovsky effect does not operate (e.g. and the off-diagonal
terms of $\mathbb{Q}$ become zero) all of the $(1/c)$ are zeroed out.
Therefore, radiative treatments which do not include the Yarkovsky effect are
potentially missing an important perturbation in the system!

Now we quantify this statement.  Because the matrices in Appendix A are all
of order unity, we can write

\begin{equation}
\left\langle \left(\frac{da}{dt}\right)_{\rm ra} \right\rangle
= \mathcal{O}\left(\frac{1}{c} \frac{AL}{4\pi mna^2} \right),
\label{avgYarki}
\end{equation}

\begin{equation}
\left\langle \left(\frac{de}{dt}\right)_{\rm ra} \right\rangle
= \mathcal{O}\left(\frac{1}{c} \frac{AL}{8\pi mna^3} \right),
\label{avgYarke}
\end{equation}

\begin{equation}
\left\langle \left(\frac{di}{dt}\right)_{\rm ra} \right\rangle
= \mathcal{O}\left(\frac{1}{c} \frac{AL}{8\pi mna^3} \right),
\end{equation}

\begin{equation}
\left\langle \left(\frac{d\Omega}{dt}\right)_{\rm ra} \right\rangle
= \mathcal{O}\left(\frac{1}{c} \frac{AL}{8\pi mna^3} \right),
\end{equation}

\begin{equation}
\left\langle \left(\frac{d\omega}{dt}\right)_{\rm ra} \right\rangle
= \mathcal{O}\left(\frac{1}{c} \frac{AL}{32\pi mna^3} \right)
.
\label{avgYarkf}
\end{equation}

\noindent{}Isolating the rate of change of eccentricity with time yields

\[
\mathcal{O}\left(\frac{1}{c} \frac{AL}{8\pi mna^3} \right)
\sim
\frac{0.08}{\rm Myr}
\left( \frac{M_{\star}}{1 M_{\odot}} \right)^{-1/2}
\left( \frac{\rho}{2 \ {\rm g/cm}^3} \right)^{-1}
\]

\begin{equation}
\ \ \ \ \ \ \ \ 
\times
\left( \frac{R}{1 \ {\rm km}} \right)^{-1}
\left( \frac{a}{5 \ {\rm au}} \right)^{-3/2}
\left( \frac{L}{10^3 L_{\odot}} \right)
.
\label{cterm}
\end{equation}

\noindent{}In other words, the Yarkovsky effect alone can fling an initially near-circular 
asteroid out of the system after about 10 Myr!  This case however assumes that changes
in the asteroid's rotation state from YORP and close encounters with the
star are neglected.  For highly eccentric orbits, both effects are likely to
cause a substantial change in the spin state of the body, which -- in turn --
could eventually eliminate the Yarkovsky effect.  Also, the interplay between
the asteroid's rotation and the stellar wind should be investigated.

In order to compare to the non-Yarkovsky term
(which includes Poynting-Robertson drag and ``radiation pressure''), note 
the following standard result (also seen in equation \ref{dedtavgY0}) but 
applied to GB stars

\[
\mathcal{O}\left(\frac{1}{c^2} \frac{5AL}{8\pi ma^2} \right)
\sim
\frac{1.8 \times 10^{-5}}{\rm Myr}
\left( \frac{\rho}{2 \ {\rm g/cm}^3} \right)^{-1}
\]

\begin{equation}
\ \ \ \ \ \ \ \ 
\times
\left( \frac{R}{1 \ {\rm km}} \right)^{-1}
\left( \frac{a}{5 \ {\rm au}} \right)^{-2}
\left( \frac{L}{10^3 L_{\odot}} \right)
,
\label{c2term}
\end{equation}

\noindent{}which is over three orders of magnitude less powerful.

Although the evolution due to time-dependent radiation is more complex 
than equations (\ref{cterm})-(\ref{c2term}) suggest (see Appendix A),
the sheer magnitude of equation (\ref{cterm}) reveals something
fundamental about the physics in these systems.  Possibilities
include (1) that asteroids in fact escape GB systems en masse,
with or without the help of mass loss, depending on the stellar wind
evolution, (2) asteroids survive but
are widely dispersed in eccentricity and inclination, (3)
the temporal variations in the individual elements of $\mathbb{Q}$ 
cancel out secular changes in the orbits, and (4) the value of $k$
is typically exceptionally low and/or the value of $Q_{\rm ref}$
is within a tiny fraction of unity so that Yarkovsky is quashed.  This 
degeneracy in interpretation
cannot be broken without adopting a detailed internal model of the target and in
particular how its incoming radiation is redistributed and how its
spin axis changes with time.  Such a model is well beyond the scope
of this paper. 

Further, and unhelpfully, we can pose arguments against all four possibilities.
Possibility (1) is unlikely because, as described in Section 1,
asteroids represent the most likely candidate for the WD pollution
that is currently observed.  Possibility (2) might be prevented from occurring
due to collisions with planets\footnote{Alternatively, distant planet-asteroid
interactions due to post-MS planet-planet scattering could repress or excite both
eccentricity and inclination \citep{verarm2005,verarm2006,rayetal2010,matetal2013}.}, 
other asteroids or smaller bodies,
depending on the architecture of the exosystem in question.  For possibility (3),
the equations in Appendix A demonstrate that even if the off-diagonal nonzero elements
of $\mathbb{Q}$ cancel one another, other combinations of elements will not cancel. 
Although the equations were
derived assuming position and velocity-independent values of $\mathbb{Q}$, the
current equations represent an accurate evolutionary picture at system snapshots
when the seasonal component is neglected.
Finally, the extreme values of $k$ and $Q_{\rm ref}$ that are required for possibility 
(4) do not conform to what we observe in the Solar system. 

We can place these results in the context of the asteroids observed
in the Solar System.  These asteroids, exposed to the relatively
weak MS luminosity of a $1M_{\odot}$ star, can experience a
significant drift in eccentricity ($\sim 0.1$) over 1 Gyr
\citep[e.g. Figs. 7-8 of][]{voketal2006}.  Because these Eros family asteroids
are located at $a \approx 3$ au and with $R > 3.5$ km, this comparison
yields good agreement with equation (\ref{cterm}).
Note that the semimajor axis drift in these asteroids over this time
period is just 0.1 au.  By estimating the semimajor
axis drift from equation (\ref{avgYarki}), we obtain

\[
\mathcal{O}\left(\frac{1}{c} \frac{AL}{4\pi mna^2} \right)
\sim
\frac{0.81 \ {\rm au}}{\rm Myr}
\left( \frac{M_{\star}}{1 M_{\odot}} \right)^{-1/2}
\left( \frac{\rho}{2 \ {\rm g/cm}^3} \right)^{-1}
\]

\begin{equation}
\ \ \ \ \ \ \ \ 
\times
\left( \frac{R}{1 \ {\rm km}} \right)^{-1}
\left( \frac{a}{5 \ {\rm au}} \right)^{-1/2}
\left( \frac{L}{10^3 L_{\odot}} \right)
.
\label{caterm}
\end{equation}

\noindent{}For a Solar-type star ($L = L_{\odot}$), this level of drift
is within the same order of magntiude as the maximum that is observed and
predicted in the Solar system \citep[e.g.][]{faretal1998,botetal2006,voketal2006}.


\subsubsection{Averaged equations with no Yarkovsky}

Recall that when Yarkovsky is not active, we are still left with
(much smaller) radiative perturbations from Poynting-Robertson drag
and radiation pressure.  We compute exactly these less-powerful ($1/c^2$) 
terms to be

\[
\left\langle 
\left(
\frac{da}{dt} 
\right)_{\rm ra}\bigg|_{\mathbb{Y}=0}
\right\rangle
= 
-
\left( \frac{1}{c^2} \right)
\frac{AL\left(Q_{\rm abs} + Q_{\rm ref}\right) \left(2 + 3 e^2\right)}
{4 \pi m a \left(1 - e^2\right)^{3/2} }
,
\]

\begin{equation}
\label{dadtavgY0}
\end{equation}

\begin{equation}
\left\langle 
\left(
\frac{de}{dt} 
\right)_{\rm ra}\bigg|_{\mathbb{Y}=0}
\right\rangle
= 
-
\left( \frac{1}{c^2} \right)
\frac{5 AL\left(Q_{\rm abs} + Q_{\rm ref}\right) e}
{8 \pi m a^2 \sqrt{1 - e^2} }
,
\label{dedtavgY0}
\end{equation}

\begin{equation}
\left\langle 
\left(
\frac{di}{dt} 
\right)_{\rm ra}\bigg|_{\mathbb{Y}=0}
\right\rangle
=
\left\langle 
\left(
\frac{d\Omega}{dt} 
\right)_{\rm ra}\bigg|_{\mathbb{Y}=0}
\right\rangle
=
0
,
\label{didtavgY0}
\end{equation}

\begin{equation}
\left\langle 
\left(
\frac{d\omega}{dt} 
\right)_{\rm ra}\bigg|_{\mathbb{Y}=0}
\right\rangle
=
\left\langle 
\left(
\frac{d\varpi}{dt} 
\right)_{\rm ra}\bigg|_{\mathbb{Y}=0}
\right\rangle
=
0
.
\label{domdtavgY0}
\end{equation}

Equations (\ref{dadtavgY0})-(\ref{dedtavgY0}), with a coefficient of
$1/c^2$, reproduce now-standard 
results from \cite{wyawhi1950}.  Their relative importance is showcased by comparing
them to equations (\ref{dadtavgYark})-(\ref{domegadtavgYark}), which
instead harbour terms with a coefficient of $1/c$.  Consequently, the Yarkovsky
effect {\it when active, dominates the target's secular evolution over other
radiative effects}.  Yarkovsky causes changes that are absent from immediately
reflected radiation.  Equations (\ref{didtavgY0})-(\ref{domdtavgY0}) are
also important because they reveal that when Yarkovsky turns on, a target
will drift out of the orbital plane that it previously occupied.

\section{Summary}

Post-main-sequence stellar radiation and winds play a decisive role
in determining the final orbital states of asteroids, pebbles and smaller 
particles.  {\it Any bodies smaller than about 1000 km will be 
significantly affected by forces other than gravity during GB evolution}. 
This paper has quantified this main conclusion through two avenues: 
a rough comparison of the instantaneous accelerations caused by mass loss, 
wind drag and radiation (Section 3) and a precise detailing of the orbital 
element equations of motions for these forces both during a single orbit and over
secular timescales (Section 4 + Appendix).

Several ancillary results are widely applicable to planetary systems at all
stages of evolution, including pre-MS and beginning MS stages.  We summarize
these results as (1) a self-consistent framework to treat Poynting-Robertson drag,
radiation pressure and the Yarkovsky effect (equation \ref{finaldvdtp}), (2) an
orbital element characterization of the Reynolds number for stellar
winds (equations \ref{ReNum}, \ref{velnor}-\ref{gasspeed}), (3) the complete
set of orbital element evolution equations for a particle dragged by gas in
the Epstein and Stokes regimes (equations \ref{unavgepsi}-\ref{finalsto}),
(4) the complete unaveraged set of equations due to radiative perturbations
assuming position and velocity-independent diurnal and seasonal Yarkovsky 
components (equations \ref{firstDeq}-\ref{lastDeq}),
and (5) the leading order of the averaged set of these equations 
(\ref{dadtavgYark}-\ref{domegadtavgYark}) when the seasonal Yarkovsky
component is negligible.  This last set, along with equations
(\ref{cterm})-(\ref{c2term}), demonstrate the potential for the Yarkovsky
effect alone to dominate the eccentricity and inclination evolution of bodies
larger than 1-10 m.

Detailed models of the time-varying stellar wind and the physical properties
of an orbiting body are required to determine its ultimate fate.  Here, we have
provided a set of machinery in which such models may be incorporated.

\section*{Acknowledgments}

We thank the referee for a thorough, probing report that has improved the
manuscript.  We also thank Holly Capelo, Alan W. Harris and Anders Johansen for 
useful discussions.  BTG and DV benefited by support by the European Union through ERC grant number 320964.
SE would like to acknowledge the support of the NEOShield project funded by the European
Union Seventh Framework Program (FP7/2007-2013) under grant agreement no. 282703
(NEOShield) as well as Paris Observatory's ESTERS (Environnement Spatial de 
la Terre : Recherche \& Surveillance) travel grants.

\appendix  
\onecolumn

\section{Explicit expressions for averaged Yarkovsky equations}

Here we write out the expressions from equations (\ref{avgYarki})-(\ref{avgYarkf}),
and do so in a form which elucidates how the leading order term ($1/c$) vanishes when
the Yarkovsky effect is turned off.  We remind the reader that an implicit assumption
in these equations is that the components of $\mathbb{Q}$ are independent of $\vec{r}$
and $\vec{v}$.

When the Yarkovsky effect plays no role in
the motion, then all of the off-diagonal terms of $\mathbb{Q}$ vanish, such
that $\mathbb{Q}_{12} = \mathbb{Q}_{21} = \mathbb{Q}_{13} = \mathbb{Q}_{31} 
= \mathbb{Q}_{23} = \mathbb{Q}_{32} = 0$.  Further, the diagonal terms become equal,
so that $\mathbb{Q}_{11} = \mathbb{Q}_{22} = \mathbb{Q}_{33}$.  Also,
in the subsequent expressions, we compress commonly-occurring eccentricity-based quantities as

\begin{equation}
\kappa \equiv -2 + e^2 + 2 \sqrt{1-e^2}
.
\label{kappaeq}
\end{equation}

\noindent{}The final expressions are

\begin{equation}
\left\langle \left(\frac{da}{dt}\right)_{\rm ra} \right\rangle
=
\left(\frac{1}{c}\right)
\frac
{AL}
{4\pi mna^2 \left(1 - e^2\right)  }
\left( \begin{array}{c} 
            \mathbb{Q}_{21} - \mathbb{Q}_{12}  \\
            \mathbb{Q}_{32} - \mathbb{Q}_{23}  \\
            \mathbb{Q}_{31} - \mathbb{Q}_{13}  \\
            \end{array}
            \right)
\cdot
\left( \begin{array}{c} 
            \cos{i}  \\
            \sin{i} \sin{\Omega}  \\
            \sin{i} \cos{\Omega}  \\
            \end{array}
            \right) 
+\mathcal{O} 
\left( \frac{1}{c^2} \frac{AL}{ma} \right),
\label{dadtavgYark}
\end{equation}

\

\

\[
\left\langle \left(\frac{de}{dt}\right)_{\rm ra} \right\rangle
=
\left(\frac{1}{c}\right)
\frac
{AL\left(1 - \sqrt{1-e^2} \right)}
{8\pi mna^3 e }
\left( \begin{array}{c} 
            \mathbb{Q}_{21} - \mathbb{Q}_{12}  \\
            \mathbb{Q}_{32} - \mathbb{Q}_{23}  \\
            \mathbb{Q}_{31} - \mathbb{Q}_{13}  \\
            \end{array}
            \right)
\cdot
\left( \begin{array}{c} 
            \cos{i}  \\
            \sin{i} \sin{\Omega}  \\
            \sin{i} \cos{\Omega}  \\
            \end{array}
            \right)
\]

\[
\ \ \ \ \
+
\left(\frac{1}{c}\right)
\frac
{AL\sqrt{1-e^2} \kappa}
{8\pi mna^3 e^3}
\left( \begin{array}{c} 
            \mathbb{Q}_{12} + \mathbb{Q}_{21}  \\
            \mathbb{Q}_{23} + \mathbb{Q}_{32}  \\
            \mathbb{Q}_{13} + \mathbb{Q}_{31}  \\
            \mathbb{Q}_{22} - \mathbb{Q}_{11}  \\
            \mathbb{Q}_{11} + \mathbb{Q}_{22} - 2\mathbb{Q}_{33} \\
            \end{array}
            \right)
\cdot
\left( \begin{array}{c} 
            \cos{i} \cos{2\omega} \cos{2\Omega} - \left(1 + \cos^2{i}\right)\cos{\Omega}\sin{2\omega}\sin{\Omega}  \\
            \sin{i} \left[ \cos{i} \cos{\Omega} \sin{2\omega} + \cos{2\omega} \sin{\Omega} \right]  \\
            \sin{i} \left[-\cos{i} \sin{\Omega} \sin{2\omega} + \cos{2\omega} \cos{\Omega} \right]  \\
            \cos{i}\cos{2\omega}\sin{2\Omega} + \left(1 + \cos^2{i} \right) \cos{\Omega} \sin{2\omega} \sin{\Omega} \\
           -\sin^2{i} \sin{\omega}\cos{\omega}  \\
            \end{array}
            \right)
\]

\begin{equation}
\ \ \ \ \
+\mathcal{O} 
\left( \frac{1}{c^2} \frac{AL}{ma^2} \right),
\label{davgYarkde}
\end{equation}

\

\

\[
\left\langle \left(\frac{di}{dt}\right)_{\rm ra} \right\rangle
=
\left(\frac{1}{c}\right)
\frac
{AL}
{8\pi mna^3 e^2\sqrt{1-e^2} }
\left[
\mathbb{Q}
\cdot
\left( \begin{array}{c} 
            \sin{i} \sin{\Omega}  \\
            -\sin{i} \cos{\Omega}  \\
            \cos{i}  \\
            \end{array}
            \right)
\right]
\cdot
\left( \begin{array}{c} 
            e^2\cos{\Omega} + \kappa \left(\cos{i}\sin{2\omega}\sin{\Omega} - \cos{2\omega} \cos{\Omega} \right)  \\
            e^2\sin{\Omega} - \kappa \left(\cos{i}\sin{2\omega}\cos{\Omega} + \cos{2\omega} \sin{\Omega} \right)  \\
            -\kappa \sin{2\omega} \sin{i}  \\
            \end{array}
            \right)
\]

\begin{equation}
\ \ \ \ \
+\mathcal{O} 
\left( \frac{1}{c^2} \frac{AL}{ma^2} \right),
\label{davgYarkdi}
\end{equation}

\

\

\[
\left\langle \left(\frac{d\Omega}{dt}\right)_{\rm ra} \right\rangle
=
\left(\frac{1}{c}\right)
\frac
{AL}
{8\pi mna^3 e^2\sqrt{1-e^2} }
\left[
\mathbb{Q}
\cdot
\left( \begin{array}{c} 
            \sin{\Omega}  \\
            -\cos{\Omega}  \\
            \cot{i}  \\
            \end{array}
            \right)
\right]
\cdot
\left( \begin{array}{c} 
            -e^2\cos{i}\sin{\Omega} - \kappa \left(\cos{i}\cos{2\omega}\sin{\Omega} + \sin{2\omega} \cos{\Omega} \right)  \\
            e^2\cos{i}\cos{\Omega} + \kappa \left(\cos{i}\cos{2\omega}\cos{\Omega} - \sin{2\omega} \sin{\Omega} \right)  \\
            e^2 \sin{i} + \kappa \cos{2\omega} \sin{i}  \\
            \end{array}
            \right)
\]

\begin{equation}
\ \ \ \ \
+\mathcal{O} 
\left( \frac{1}{c^2} \frac{AL}{ma^2} \right),
\end{equation}

\begin{equation}
\left\langle \left(\frac{d\omega}{dt}\right)_{\rm ra} \right\rangle
=
\left(\frac{1}{c}\right)
\frac
{AL}
{32\pi mna^3 e^4\sqrt{1-e^2} }
\cdot
\left( \begin{array}{c} 
            \mathbb{Q}_{12} - \mathbb{Q}_{21}  \\
            \mathbb{Q}_{23} - \mathbb{Q}_{32}  \\
            \mathbb{Q}_{13} - \mathbb{Q}_{31}  \\
            \mathbb{Q}_{12} + \mathbb{Q}_{21}  \\
            \mathbb{Q}_{23} + \mathbb{Q}_{32}  \\
            \mathbb{Q}_{13} + \mathbb{Q}_{31}  \\
            \mathbb{Q}_{11} + \mathbb{Q}_{22}  \\
            \mathbb{Q}_{11} - \mathbb{Q}_{22}  \\
            \mathbb{Q}_{11} + \mathbb{Q}_{22} - 2\mathbb{Q}_{33} \\
            \end{array}
\right)
\cdot
\left( \begin{array}{c} 
            \Upsilon_1  \\
            \Upsilon_2  \\
            \Upsilon_3  \\
            \Upsilon_4  \\
            \Upsilon_5  \\
            \Upsilon_6  \\
            \Upsilon_7  \\
            \Upsilon_8  \\
            \Upsilon_9  \\
            \end{array}
\right)
+
\mathcal{O} 
\left( \frac{1}{c^2} \frac{AL}{ma^2} \right)
\label{domegadtavgYark}
\end{equation}

\noindent{}where the auxiliary $\Upsilon$ variables are long explicit functions of $(e,i,\Omega,\omega)$.  We do
not write out these functions here but are happy to provide them to interested readers.

\twocolumn

\label{lastpage}

\end{document}